\documentclass[conference]{IEEEtran}
\usepackage{times}

\usepackage{graphics} 
\usepackage{epsfig} 
\graphicspath{{picture/}}
\usepackage{amsmath} 
\usepackage{amssymb}  
\usepackage{color}
\usepackage{amsthm}
\usepackage{algorithm}
\usepackage{tikz} 
\usepackage{makecell}
\usepackage{multirow}
\usepackage{algpseudocode}
\usepackage{enumerate}

\usepackage{stfloats}
\usepackage{caption}
\usepackage{subcaption}

\usepackage[numbers]{natbib}
\usepackage{multicol}

\usepackage{hyperref}
\usepackage{cleveref}

\theoremstyle{definition}

\newtheorem{remark}{Remark}

\newtheorem{theorem}{Theorem}
\newtheorem{corollary}{Corollary}

\begin{document}

\title{Safe Multi-Agent Navigation via Constrained HJB-Informed Learning}


\author{
\authorblockN{
Fenglan Wang,
Xinguo Shu,
Lei He, and
Lin Zhao\textsuperscript{*}
}
\authorblockA{
Department of Electrical and Computer Engineering, 
National University of Singapore, 
Singapore 117583\\
Email: wfenglan@nus.edu.sg (F. Wang), e1352650@u.nus.edu (X. Shu),\\
lei.he@nus.edu.sg (L. He), zhaolin@nus.edu.sg (L. Zhao)
}
\thanks{Corresponding author: Lin Zhao.}
}


%

\twocolumn[{
  \begin{@twocolumnfalse}
    \maketitle 
    
    \vspace{-1.6em} 
    \centering
    \includegraphics[width=1.0\textwidth]{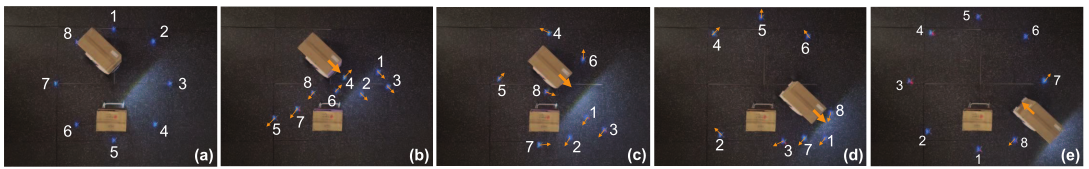}
    
    \captionof{figure}{
    \textbf{Antipodal position swapping of eight Crazyflie miniature drones using the proposed HJB-GNN navigation policy under limited sensing range in an unseen environment containing static and dynamic obstacles.}
HJB-GNN is a Hamilton-Jacobi-Bellman (HJB) equation-based learning framework for distributed infinite-horizon safe optimal navigation using graph neural networks (GNN). It enables a principled and adaptive trade-off between goal reaching and collision-avoidance through graph-dependent Lagrange multipliers derived from Karush-Kuhn-Tucker (KKT) optimality conditions. (a) Starting positions. (b)-(e) Temporal snapshots illustrating the evolution of the drone and obstacle trajectories. Orange arrows indicate the velocity directions of the drones and a large, randomly moving dynamic obstacle. 
 }
    \label{fig-mas}
    \vspace{0.4em} 
  \end{@twocolumnfalse}
}]

\begingroup
\renewcommand{\thefootnote}{*}
\footnotetext{Corresponding author}
\endgroup

\begin{abstract}
Multi-agent navigation in unknown and cluttered environments has broad applications, yet remains fundamentally challenging. In particular, dense agent-agent and agent-obstacle reactive interactions can exacerbate the inherent competition between collision-avoidance constraints and goal-reaching objectives. Most existing approaches mitigate this by applying per-step safety filtering on top of a predefined goal-reaching controller or by designing heuristic loss functions that penalizes safety constraints violation gradient. 
While effective in sparse environments, these methods still suffer from overly-conservative behaviors 
when interactions become dense.
To overcome these limitations, 
we propose HJB-GNN, a Hamilton-Jacobi-Bellman (HJB)-based learning framework that jointly learns a graph neural network (GNN)-parameterized control barrier function for explicit safety enforcement, a distributed GNN-based navigation policy, and a value function that induces goal-reaching behavior. By exploiting the analytical solution of the constrained HJB equation, the proposed method derives graph-dependent Lagrange multipliers that adaptively balance collision-avoidance and goal-reaching across diverse multi-agent navigation scenarios.
Moreover, HJB-GNN supports centralized training with distributed deployment. Extensive simulations and real-world experiments with Crazyflie drone swarms demonstrate its superior safety and goal-reaching performance, as well as strong scalability and generalizability to large-scale teams operating in previously unseen, dense environments. \footnote{Project website: \href{https://github.com/hublan24/HJB-GNN}{https://github.com/hublan24/HJB-GNN}.}

\end{abstract}

\IEEEpeerreviewmaketitle

\section{Introduction}
Multi-agent navigation is a core capability underpinning a wide range of real-world robotic systems, including autonomous mobile robots in automated warehouses~\cite{Wang2023SMC,zhang2021safe,Mestres2024ral}, drone swarms for search-and-rescue operations~\cite{zhang2025tro,Vinod2024tcst,Choi2025Resolving,Safaoui2024tro}, and unmanned surface vehicles for coordinated patrol missions~\cite{huang2024tiv}.
Due to limited onboard computation, sensing range, and communication bandwidth, practical deployments typically rely on distributed navigation controllers that use only local neighborhood information~\cite{wu2025smc,yan2025tnnls,Mrdjan2024cst,farivarnejad2024local,Wang2017tro,Choi2025Resolving}.
However, in dense, cluttered, and unknown environments, frequent reactive interactions among agents and obstacles substantially intensify the conflict between collision-avoidance constraints and goal-reaching objectives, posing significant challenges for reliable and efficient navigation.

Recent works have attempted to mitigate the conflict between collision-avoidance and goal-reaching objectives through safety filtering. In particular, several approaches~\cite{wu2025smc,Dimitratrust,Dimarogonasa,yan2025tnnls,Mrdjan2024cst,Mestres2024ral,Wang2017tro} employed quadratic programming (QP)-based safety filters, where a predefined nominal goal-reaching controller was often projected onto control barrier function (CBF)-based collision avoidance constraints to ensure safety. 
However, these works are limited in \textit{known} environments where manually specifying CBF is tractable. For multi-agent systems (MASs) navigating in \textit{unknown} environments under limited sensing range, a line of recent work has focused on learning CBFs directly from data~\cite{zhang2025tro,gaoprovably}. Notably,~\cite{zhang2025tro} proposed graph neural networks (GNNs)-parameterized CBFs (graph CBF or GCBF) for MAS navigation, which naturally adapts to dynamically changing interaction graphs. A distributed GNN-based navigation policy was then learned by imitating the corresponding safety-filtered controller. 
While it (hereafter referred to as QP-GCBF) demonstrates promising empirical performance, its effectiveness is fundamentally constrained by its reliance on a predefined, fixed nominal controller. To alleviate this limitation,~\cite{gaoprovably} incorporated control Lyapunov function (CLF) constraints into the QP-based framework, thereby avoiding explicit specification of a nominal controller. 
Nevertheless, both~\cite{zhang2025tro} and~\cite{gaoprovably} provided only one-step-ahead safety guarantees. The induced control policies are inherently myopic and can drive agents excessively close to obstacles or other agents, often leading to overly-conservative behavior, deadlocks, or even collisions in dense multi-agent scenarios.

Another line of more recent works have further proposed MAS navigation solutions in a constrained infinite-horizon optimal control formulation~\cite{zhang2025discrete,zhang2025defmarl}. In particular, \cite{zhang2025discrete} proposed learning discrete-time graph CBF-constrained navigation policies using proximal policy optimization \cite{ppo} in a model-free constrained reinforcement learning framework. They mitigate the conflicts between safety and goal-reaching (task) in the loss design through an approximate gradient projection mechanism, where task gradients are projected to be orthogonal to the safety constraints gradients. However, their safety-performance trade-off is still regulated by heuristic scheduling of penalty weights, which tends to induce overly-conservative behaviors in dense and cluttered environments due to complex interaction constraints. 
Furthermore, to enable stable training, \cite{zhang2025defmarl} developed a distributed learning algorithm based on an epigraph reformulation, in which an auxiliary variable is minimized as an upper bound on the task cost. However, this reformulation achieves collision-avoidance and goal-reaching coordination indirectly through the evolution of the auxiliary variable, which can limit the agent’s responsiveness to rapidly changing interactions in dense environments and easily result in deadlocks or collisions.

To tackle these limitations, we propose a novel infinite-horizon CBF-constrained optimal control formulation for safe multi-agent navigation in unknown environments with limited sensing radii.
We formally analyze the resulting constrained optimal control problem through the lens of Hamilton-Jacobi-Bellman (HJB) equations and Lagrange duality, and develop a physics-informed learning framework that leverages GNNs to approximate the solution.
Unlike existing approaches, which do not explicitly optimize the trade-off between collision-avoidance and goal-reaching, our method exploits the analytical structure of constrained HJB formulation to enable principled, state- and graph-dependent coordination between these competing objectives. As a result, the learned distributed policy effectively handles dense agent-agent and agent-obstacle interactions, substantially reducing deadlocks and collisions in practice (see Fig.~\ref{fig-mas} for a challenging scenario).
More specifically, our main contributions are as follows:

(i) 
We establish an infinite-horizon safe optimal control framework for MASs using graph CBF. The proposed framework introduces an adaptive trade-off mechanism for safety and goal-reaching performance via graph-dependent Lagrange multipliers. Such graph-dependent multipliers, derived from Karush-Kuhn-Tucker (KKT) optimality conditions, effectively coordinate collision-avoidance and goal-reaching control to adapt to sensing-based changing interaction graph, thereby reducing the conservative behaviors induced by manually tuned static parameter, heuristic gradient projection, or auxiliary variable-based trade-off designs~\cite{Dawood2025RAL,zhang2023neural,Yang2024tac,zhang2025discrete,zhang2025defmarl}.

(ii) 
We develop a novel HJB-GNN learning algorithm that enables end-to-end joint synthesis of a value function, a graph CBF, and a distributed policy using GNNs, guided by the analytical safe optimal solution structure of constrained HJB equations. 
The proposed HJB-GNN algorithm 
eliminates the reliance on short-horizon QP solvers and predefined nominal controllers commonly used in existing safety control methods \cite{zhang2025tro, Wang2017tro,gaoprovably,DawsonLearning}.

(iii) 
We adopt a centralized training and a distributed deployment paradigm for the HJB-GNN algorithm. Policy trained on systems with only eight agents generalize to scenarios involving hundreds and even thousands of agents, and scale effectively to previously unseen, dense environments. Extensive simulations demonstrate significantly improved safety, scalability, and generalizability compared with the state-of-the-art QP-GCBF method \cite{zhang2025tro}. 
We validate the applicability and robustness of the HJB-GNN through hardware experiments using Crazyflie drone swarms navigating in complex obstacle environments. \footnote{Simulation and  experiment videos can be found at: \href{https://nus-core.github.io/assets/standalone/HJB-GNN/index.html}{https://nus-core.github.io/assets/standalone/HJB-GNN/index.html}.}

The remainder of the paper is organized as follows. Section~\ref{Section2} reviews preliminaries. Section~\ref{Section3} formulates the constrained HJB problem with a graph CBF and analyzes its solution. Section~\ref{Section4} presents the HJB-GNN learning framework. Simulations and hardware experiments are given in Sections~\ref{Section5}. Section~\ref{Section7} concludes the paper. 

\emph{Notations:} 
$\nabla_x W$ represents the derivative of the function $W$ with respect to $x$ and is interpreted as a row vector, and $\nabla_x^T W$ represents the transpose of $\nabla_x W$. 
A continuous function $\alpha:[-b,a)\to \mathbb{R}$ with $a,b>0$, is of extended class $\mathcal{K}$ ($\alpha\in\mathcal{K}_e$) if $\alpha$ satisfies $\alpha(0)=0$ and is strictly increasing. We denote $(a,b)=[a^T, b^T]^T$ for column vectors $a$ and $b$.

\section{Preliminaries}\label{Section2}

\subsection{System Dynamics}\label{Section2.1}
We aim to study the problem of distributed safe multi-agent navigation, where each agent $i\in V_a:=\{1,2,...,N\}$, is required to reach its goal position while 
avoiding collisions with other agents and static obstacles. The agents are governed by the following dynamics: 
\begin{align}\label{mas}
\dot{\bold{x}}_i=\bold{f}(\bold{x}_i)+\bold{g}(\bold{x}_i)\bold{u}_i,\ i=1,2,...,N,
\end{align}
where for the $i$-th agent, $\bold{x}_i\in \mathcal{X}\subset\mathbb{R}^n$ and $\bold{u}_i\in \mathcal{U}\subset\mathbb{R}^m$ denote the state and control input, respectively, $\mathcal{X}$ and $\mathcal{U}$ are two compact sets, and $\bold{f}:\mathcal{X}\to\mathbb{R}^{n}$, $\bold{g}:\mathcal{X}\to\mathbb{R}^{n\times m}$ are two locally Lipschitz continuous functions. Denote $\bold{\bar{x}}:=(\bold{x}_1,\bold{x}_2,...,\bold{x}_N)\in \mathcal{X}^N$ and $\bold{\bar{u}}:=(\bold{u}_1,\bold{u}_2,...,\bold{u}_N)^T\in \mathcal{U}^N$ as the state vector and the control input vector of the MAS (\ref{mas}), respectively.

Each agent has a \emph{limited} sensing radius $R_a\in\mathbb{R}_{>0}$ and thus communicates with other agents and observes the obstacle environment only within its sensing range. The interactions among agents and obstacles are represented by a direct graph $G=(V, E)$, where $V=V_a\cup V_o$ is the set of nodes which contain the set of agent nodes $V_a$ and the set of obstacle nodes $V_o$, and $E\subset\{(i,j)|i\in V_a,j\in V\}$ denotes the set of edges. An edge $(i,j)\in E$  implies that the agent $i$ communicates with other agent $j$ or observes an obstacle $j$. 
The topology graph $G$ and its edge set $E$ are allowed to vary arbitrarily, without requiring the fixed number of neighbors assumed in \cite{zhang2025towardtro,saravanos2023distributed}.

\subsection{Safety of Multi-Agent Systems}\label{Section2.2}
Denote a position space by $\mathbb{P}\subset\mathbb{R}^{n_d}$, where $n_d\in\{2,3\}$ represents the spatial dimensions of environments. In the MAS (\ref{mas}), assume that the first $n_d$ entries of the state vector $\bold{x}_i$ correspond to the position $\bold{p}_i\in\mathbb{P}$ of agent $i\in V_a$. We denote the vector of relative positions information observed by each agent $i\in V_a$ from all neighboring obstacles within its sensing radius as  
$\bold{O}_i:=(\bold{O}_{i1},\bold{O}_{i2},...,\bold{O}_{in_{oi}})$ with $n_{oi}\in\mathbb{N}$. 
To ensure collision-avoidance among agent-agent and agent-obstacle, 
denote the overall safe set $\mathcal{S}\subset \mathcal{X}^{N}$ in the MAS (\ref{mas}) as: 
\begin{subequations}\label{S}
\begin{align}
 \mathcal{S}:=&\cap_{i=1}^{N}\mathcal{S}_i,\\
 \mathcal{S}_i:=&\Big\{\bold{\bar{x}}\in \mathcal{X}^N\Big|\big\|
\bold{O}_{ij}\big\|>r,\, \forall j=1,2,...,n_{oi},\nonumber\\ 
&\qquad\qquad\min_{i,j\in V_a,i\neq j}\big\|\bold{p}_{i}-\bold{p}_j\big\|>2r\Big\},  
\end{align}
\end{subequations}
where $r$ represents the radius of a minimal circle enclosing each agent's physical body with $2r<R_a$. 
The MAS (\ref{mas}) is safe with respect to the set $\mathcal{S}$ if for any initial state $\bold{\bar{x}}(t_0)\in \mathcal{S}$, there exists a controller $\bold{\bar{u}}$, such that $\bold{\bar{x}}(t)\in \mathcal{S}$, $\forall t\ge t_0$.

We employ a graph CBF to deal with safety constraints for the MAS~\eqref{mas} with arbitrarily varying neighborhoods inspired by~\cite{zhang2025tro}. For each agent $i$, let $\mathcal{N}_{i}$ denote the neighbor set of agent $i$, that is,
\begin{align}\label{N_i_tilde}
\mathcal{N}_{i}:=\big\{j\in V \big|\|\bold{p}_i-\bold{p}_j\|\leq R_a,\|\bold{O}_{ij}\|\leq R_a\big\}.
\end{align} 
An augmented neighborhood set $\hat{\mathcal N}_i$ consists of $M$ \textit{closest} agents and obstacles if $|\mathcal{N}_{i}| > M$, and otherwise set $\hat{\mathcal{N}}_i = \mathcal{N}_{i}$. The corresponding augmented neighborhood state $\bar{\boldsymbol x}_{\hat{\mathcal N}_i}\in\mathcal X^M$ is used as the input to the graph CBF by concatenating the states of agent $i$ and its neighbors, where constant-padding is applied to maintain a fixed dimension when the size of $\hat{\mathcal{N}}_i$ is less than $M$. $\check{\mathcal{N}}_i \subseteq \hat{\mathcal{N}}_i$ denotes the set of neighbors strictly within $R_a$. $\hat{\mathcal{N}}_i^a$ and $\check{\mathcal{N}}_i^a$ denote the subsets of agents within $\hat{\mathcal{N}}_i$ and $\check{\mathcal{N}}_i$, respectively. Then, the $C^1$ graph CBF $h: \mathcal{X}^M\to\mathbb{R}$ is designed such that for all $\bold{\bar{x}}\in \mathcal{X}^N$ with $N\ge M$, 
\begin{align}\label{CBF_N_R}
\dot{h}(\bold{\bar{x}}_{\hat{\mathcal{N}}_i})\!=&\sum_{l\in \hat{\mathcal{N}}_i^a}\!\!\nabla_{\bold{x}_l}h(\bold{\bar{x}}_{\hat{\mathcal{N}}_i})(\bold{f}(\bold{x}_l)\!+\!\bold{g}(\bold{x}_l)\bold{u}_l)\nonumber\\
\!\ge& -\alpha(h(\bold{\bar{x}}_{\hat{\mathcal{N}}_i})), 
\end{align}
where $\alpha\in\mathcal{K}_e$, $\bold{u}_l=\pi_l(\bold{\bar{x}}_{\mathcal{\hat{N}}_l})$ with $\pi_i: \mathcal{X}^M\to \mathcal{U}$, and two conditions are satisfied: $h(\bold{\bar{x}}_{\hat{\mathcal{N}}_i})=h(\bold{\bar{x}}_{\check{\mathcal{N}}_i})$ and 
$\nabla_{\bold{x}_l}h(\bold{\bar{x}}_{\hat{\mathcal{N}}_i})=0, \ \forall l\in \hat{\mathcal{N}}_i^a\setminus\check{\mathcal{N}}_i^a$.

Denote the zero-superlevel set of $h$ by $\mathcal{C}:=\{\bold{\tilde{x}}\in \mathcal{X}^M\, |\, h(\bold{\tilde{x}})\ge0\}.$ 
Let $\mathcal{H}_{N,i}$ be the set of states $\bold{\bar{x}}$ of MAS where the neighborhood state $\bold{\bar{x}}_{\hat{\mathcal{N}}_i}$ of agent $i$ is in set $\mathcal{C}$, that is, $\mathcal{H}_{N,i}:=\{\bold{\bar{x}}\in \mathcal{X}^N\, |\, \bold{\bar{x}}_{\hat{\mathcal{N}}_i}\in \mathcal{C} \}.$ 
Denote their intersection as $\mathcal{H}_{N}:=\bigcap_{i=1}^N \mathcal{H}_{N,i}.$
Clearly, $\mathcal{H}_{N,i}\subseteq \mathcal{S}_{i}$ for all $i\in V_a$ implies $\mathcal{H}_{N}\subseteq \mathcal{S}$. Without loss of generality, assume that there exists a small constant $\varrho\in\mathbb{R}_{>0}$, such that $\|\nabla_{\bold{x}_i}h(\bold{\bar{x}}_{\hat{\mathcal{N}}_i})\bold{g}(\bold{x}_i)\|\ge \varrho$ for all $\bold{\bar{x}}_{\hat{\mathcal{N}}_i}\in \mathcal{X}^M$. It is a regularity condition which makes that the CBF condition~\eqref{CBF_N_R} is effective on the system control. It can be easily satisfied by a generic GNN.

\section{Safe Multi-Agent Navigation via Infinite-Horizon Constrained Optimization}\label{Section3}

In this section, we formulate an infinite-horizon safe optimal control framework based on a graph CBF. Our framework derives a principled adaptive trade-off mechanism that balances safety and goal-reaching performance via graph-dependent Lagrange multipliers. Crucially, this formulation characterizes the analytical structure of safe optimal controller, providing a rigorous theoretical foundation for the HJB-GNN learning algorithm developed in Section~\ref{Section4}.

\subsection{Safe Optimal Control Formulation}
To address the safe multi-agent navigation problem without requiring predefined task controllers used in \cite{zhang2025tro,zhang2023neural,AlbertOneFilter}, we unify safety and goal-reaching objectives within a single infinite-horizon constrained optimal control problem:  
\begin{subequations}\label{cop}
\begin{align}
\min_{\bold{u}_i\in \mathcal{U}}&\quad 
\int_{t_0}^{\infty}r_i(\bold{e}_i(\tau),\bold{u}_i(\tau))d\tau\label{cop1}\\
\ {\rm s.t.}&\quad \eqref{mas},\,\,\eqref{CBF_N_R},\, \,
\bold{\bar{x}}(t_0)\in \mathcal{H}_{N}, \ \forall\, t\ge t_0,
\end{align}
\end{subequations}
 for all $i\in V_a$, where for $\bold{e}_i\in  \mathcal{X}$, the first $n_d$ elements denotes the goal-reaching error $\bold{p}_i-\bold{p}_i^{g}$ and the remaining $n-n_d$ entries are zero-padded, $\bold{p}_i^g \in \mathbb{P}$ is the goal position, and the immediate cost is $r_i(\bold{e}_i,\bold{u}_i):=\bold{e}_i^T\bold{Q}\bold{e}_i+\bold{u}_i^T\bold{R}\bold{u}_i$, $\bold{Q} \in \mathbb{R}^{n \times n}$ and $\bold{R} \in \mathbb{R}^{m \times m}$ are two symmetric, positive definite matrices.

\subsection{Safe Optimal Controllers Synthesis via Constrained HJB}
We reformulate the original problem (\ref{cop}) as a graph CBF-constrained HJB equation, whose solution is characterized  utilizing the KKT optimality conditions with graph-dependent Lagrange multipliers.

Denote a set of admissible policies by $\mathcal{U}_a$ as in \cite{vamvoudakis2010online} where for each $\mu\in \mathcal{U}_a$, the corresponding control input additionally needs to remain within the set $\mathcal{U}$. 
For an admissible policy $\mathbf{u}_i \in \mathcal{U}_a$, an infinite-horizon value function is defined by $V_{i}(\bold{e}_i(t)):=\int_{t}^{\infty}r_i(\bold{e}_i(\tau),\bold{u}_i(\tau)){\rm d}\tau$. The corresponding \textit{Hamiltonian} is described as:
\begin{align}\label{optimal-H}
&H_i(\bold{e}_i,\bold{u}_i,\nabla_{\bold{e}_i} V_i(\bold{e}_i))\nonumber\\
=~&\nabla_{\bold{e}_i} V_i(\bold{e}_i)\big(\bold{f}(\bold{x}_i)+\bold{g}(\bold{x}_i)\bold{u}_i\big)+r_{i}(\bold{e}_i,\bold{u}_i).
\end{align} 
Describe the optimal value function $V_i^*:\mathbb{R}^n\to\mathbb{R}$ by $V_{i}^*(\bold{e}_i(t))=\min_{\bold{u}_i\in \mathcal{U}_a}\int_{t}^{\infty}r_i(\bold{e}_i(\tau),\bold{u}_i(\tau))d\tau$. 
According to Bellman’s principle of optimality \cite{lewisoptimal}, the optimal value function satisfies the following HJB equation
\begin{align}\label{optima-lag}
0=~&\min_{\bold{u}_i\in \mathcal{U}_a}H_i(\bold{e}_i,\bold{u}_i,\nabla_{\bold{e}_i} V_i^*(\bold{e}_i))\nonumber\\
=~&H_i(\bold{e}_i,\bold{u}_i^*,\nabla_{\bold{e}_i} V_i^*(\bold{e}_i)),
\end{align}   
where $\bold{u}_i^*$ is the optimal solution of \eqref{cop}.

To address the problem \eqref{optima-lag}, it is transformed into finding a control input $\bold{u}_i\in \mathcal{U}_a$ that minimizes the Hamiltonian function, subject to the graph CBF constraint in \eqref{CBF_N_R}, for each agent $i$. The resulting constrained optimization problem is described by 
\begin{subequations}\label{const-hjb}
\begin{align}
\min_{\bold{u}_i\in \mathcal{U}_a}& \quad H_i(\bold{e}_i,\bold{u}_i,\nabla_{\bold{e}_i} V_i^*(\bold{e}_i))
,\label{const-hjb1}\\
 \text{s.t.} & \quad \dot{h}(\bold{\bar{x}}_{\hat{\mathcal{N}}_i})+\alpha(h(\bold{\bar{x}}_{\hat{\mathcal{N}}_i}))\ge 0.
 \label{const-hjb2}
\end{align}   
\end{subequations}
It is observed that \eqref{const-hjb} and
\eqref{cop} are point-wise equivalent. To address the constrained optimization problem \eqref{const-hjb} and consequently \eqref{cop}, 
we note that both the objective function~\eqref{const-hjb1} and the safety constraint~\eqref{const-hjb2} are convex with respect to the
decision variable $\bold{u}_i$. Thus we can characterize the safe optimal controller $\bold{u}_{i}^*$ by applying KKT optimality conditions which are both necessary and
sufficient for optimality. We have  
\begin{subequations}\label{kkt}
\begin{align}
\nabla_{\bold{u}_{i}^*}H_i(\bold{e}_i,\bold{u}_{i}^*,\nabla_{\bold{e}_i}\! V_i^*(\bold{e}_i))\!-\!\lambda^*_i\nabla_{\bold{x}_i}\!h(\bold{\bar{x}}_{\hat{\mathcal{N}}_i})\bold{g}(\bold{x}_i)
=\ &0,\label{kkt1}\\
\lambda^*_i\big(\dot{h}(\bold{\bar{x}}_{\hat{\mathcal{N}}_i})+\alpha(h(\bold{\bar{x}}_{\hat{\mathcal{N}}_i}))
\big)
=\ &0,\label{kkt2}\\
\dot{h}(\bold{\bar{x}}_{\hat{\mathcal{N}}_i})+\alpha(h(\bold{\bar{x}}_{\hat{\mathcal{N}}_i}))\ge\ & 0
,\label{kkt3}\\
\lambda^*_i\ge\ &0,\label{kkt4}
\end{align}   
\end{subequations}
where $\nabla_{\bold{u}_{i}^*}H_i(\bold{e}_i,\bold{u}_{i}^*,\nabla_{\bold{e}_i}\! V_i^*(\bold{e}_i))\!:=\!\nabla_{\bold{e}_i} \!V_{i}^{*}(\bold{e}_i)\bold{g}(\bold{x}_i)+2\bold{u}_{i}^{*T}\bold{R}$. 
We derive the analytical form of safe optimal controller by solving \eqref{kkt1}:
\begin{align}\label{us}
\bold{u}_{i}^*=-\frac{1}{2}\bold{R}^{-1}\bold{g}^T(\bold{x}_i)\big(\underbrace{\nabla_{\bold{e}_i}^T V_{i}^*(\bold{e}_i)}_{\text{Goal-reaching}} -\lambda^*_i\underbrace{\nabla_{\bold{x}_i}^T h(\bold{\bar{x}}_{\hat{\mathcal{N}}_i})}_{\text{Safety}}\big).
\end{align}

Substituting (\ref{us}) into the complementary slackness condition (\ref{kkt2}) yields:
\begin{align}
&\lambda^*_i\big(\Lambda_{i}+\lambda^*_i\omega_{i}\big)=0,\ \forall i\in V_a,\label{lagrange1}
\end{align}
where 
$\Lambda_{i}:=-\frac{1}{2}\nabla_{\bold{x}_i} h(\bold{\bar{x}}_{\hat{\mathcal{N}}_i})\bold{g}(\bold{x}_i)\bold{R}^{-1}\bold{g}^T(\bold{x}_i)\nabla_{\bold{e}_i}^T V_{i}^*(\bold{e}_i)+\nabla_{\bold{x}_i}h(\bold{\bar{x}}_{\hat{\mathcal{N}}_i})
\bold{f}(\bold{x}_i)\!+\!\sum_{j\in \hat{\mathcal{N}}_i^a,j\neq i}\Lambda_{ij}^{e}\!+\!\alpha(h(\bold{\bar{x}}_{\hat{\mathcal{N}}_i}))
$, $\Lambda_{ij}^{e}:=\nabla_{\bold{x}_j}h(\bold{\bar{x}}_{\hat{\mathcal{N}}_i})\big(\bold{f}(\bold{x}_j)+ \bold{g}(\bold{x}_j)\bold{u}_{j}\big), j\in \hat{\mathcal{N}}_i^a, j\neq i$, $\omega_{i}:=\frac{1}{2}\nabla_{\bold{x}_i} h(\bold{\bar{x}}_{\hat{\mathcal{N}}_i})\bold{g}(\bold{x}_i)\bold{R}^{-1}\bold{g}^T(\bold{x}_i)\nabla_{\bold{x}_i}^Th(\bold{\bar{x}}_{\hat{\mathcal{N}}_i})$. 
Given $\|\nabla_{\bold{x}_i}h(\bold{\bar{x}}_{\hat{\mathcal{N}}_i})\bold{g}(\bold{x}_i)\|\ge \varrho$ for all $\bold{\bar{x}}_{\hat{\mathcal{N}}_i}\in \mathcal{X}^M$, then $\omega_{i}>0$. Under \eqref{kkt3} and \eqref{kkt4}, solving (\ref{lagrange1}) provides the graph-dependent Lagrange multiplier: 
\begin{align}\label{lag}
\lambda^*_i=\begin{cases}
-\frac{\Lambda_{i}}{\omega_{i}},&  {\rm if}\  \Lambda_{i}<0,\\
0,& {\rm if}\ \Lambda_{i}\ge0.
\end{cases}
\end{align}

\textbf{Adaptive trade-off mechanism:} The safe optimal control \eqref{us} effectively integrates \textit{goal-reaching control component} ($\nabla_{\bold{e}_i}^T V_{i}^*(\bold{e}_i)$)
and \textit{safety control component} ($\nabla_{\bold{x}_i}^T h(\bold{\bar{x}}_{\hat{\mathcal{N}}_i})$) via the Lagrange multiplier $\lambda_i^*$. The multiplier explicitly depends on the local interaction topology graph $\hat{\mathcal{N}}_i$ of agent $i$, since the set of active safety constraints is generated by its neighbor nodes.  Such graph-dependent Lagrange multiplier serves as an adaptive trade-off factor that enables adaptive coordination between safety and goal-reaching control according to real-time changing interaction topology graph. Specifically, the term $\Lambda_i$ characterizes the safety margin only under the goal-reaching control component. $\Lambda_i \ge 0$ indicates that the goal-reaching control naturally satisfies the CBF constraint \eqref{kkt3}, and the multiplier $\lambda_i^*$ vanishes. $\Lambda_i < 0$ indicates that the goal-reaching control threatens safety, and $\lambda_i^*$ automatically activates to its optimal value $-\Lambda_i/\omega_i$. This multiply automatically increases to strengthen repulsive effects to maintain safety when neighbors are very close; and similarly, the multiply decreases accordingly to relax the strength of the safety control and prioritizes goal-reaching control when neighbor density decreases.  
Consequently, the proposed approach avoids the conservative trade-offs inherent in existing methods with manually tuned static parameter, heuristic gradient projection, or auxiliary variable-based trade-off designs~\cite{Dawood2025RAL,zhang2023neural,Yang2024tac,zhang2025discrete,zhang2025defmarl}, thereby reducing deadlocks and collisions in dense environments.

 \begin{figure}[htpb]
 \centering
\includegraphics[scale=2.05]{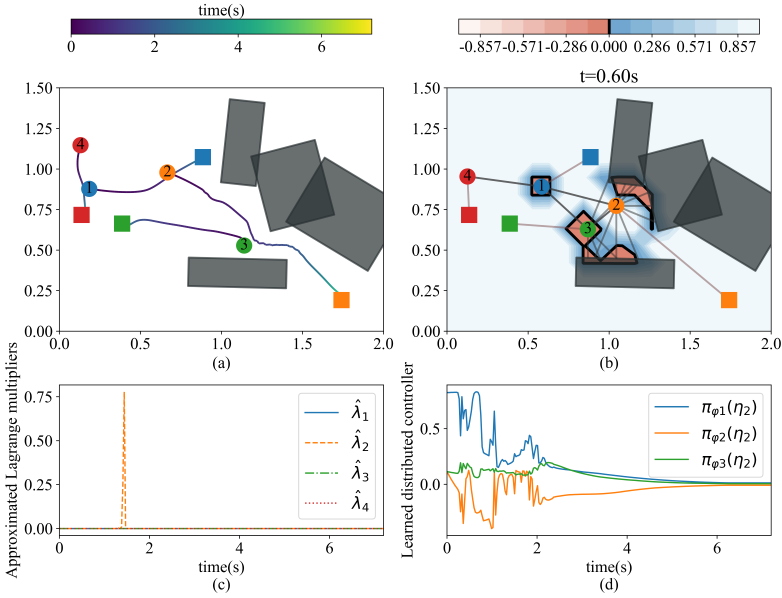}
\caption{
Adaptive trade-off between safety and goal-reaching performance for unmanned surface vessel model of \cite{namerikawa2024equivalence} in cluttered, dense environment with $2\mathrm{m} \times 1.5\mathrm{m}$. 
(a) Four agents safely navigate to their respective goals. Numbered circles denote initial positions of agents, squares with the same color as agents are their goals, and gray rectangles are static obstacles. (b) Learned graph CBF contour of agent 2 at $t=0.6{\rm s}$, where the zero-level set of learned CBF identifies the safety boundaries. (c) Approximated Lagrange multipliers $\hat{\lambda}_i, i=1,2,3,4$. (d) Learned distributed controller of agent 2. Detailed simulation setup is illustrated in Appendix\ref{App-simulation-setup}. 
}\label{fig.3}
 \end{figure}

As shown in Fig.~\ref{fig.3},  
the evolution of $\hat{\lambda}_i$ in Fig.~\ref{fig.3}c validates the adaptive trade-off: a non-zero $\hat{\lambda}_2$ emerges when agent 2 encounters a potential collision to strength safety enforcement, while other multipliers remain at zero to preserve the unconstrained navigation optimality. It demonstrates that the adaptive trade-off mechanism enables effective safe navigation in cluttered, dense environment.

The following theorem certifies both the safety and goal-reaching properties of the safe optimal controller (\ref{us}).

\begin{theorem}\label{the1}
Consider an arbitrarily sized nonlinear MAS (\ref{mas}). Suppose that there exist the optimal value function $V_{i}^*(\bold{e}_i)$ of (\ref{cop}) being $C^1$ and positive definite, and a graph CBF $h(\bold{\bar{x}}_{\hat{\mathcal{N}}_i})$ satisfying $\mathcal{H}_{N,i}\subseteq \mathcal{S}_{i}$, $ i\in V_a$ and $\|\nabla_{\bold{x}_i}h(\bold{\bar{x}}_{\hat{\mathcal{N}}_i})\bold{g}(\bold{x}_i)\|\ge \varrho$ with $\varrho\in\mathbb{R}_{>0}$ for all $\bold{\bar{x}}_{\hat{\mathcal{N}}_i}\in \mathcal{X}^M$.
Then, for any initial condition $\bold{\bar{x}}(t_0)\in \mathcal{H}_{N}$, the safe optimal controllers (\ref{us}) with the graph-dependent Lagrange multipliers (\ref{lag}) for all $i\in V_a$ guarantee that (i) the system state remains in the safe set, i.e., 
$
\bold{\bar{x}}(t) \in \mathcal{S}$ for all $ t\ge t_0,
$
and (ii) the goal-reaching error $\bold{e}_i(t)$ of each agent $i$ asymptotically converges to the zero.
\end{theorem}

\begin{IEEEproof}
See the Appendix\ref{App-proof-the}.
\end{IEEEproof}

\begin{remark}\label{rem1}
In Theorem~\ref{the1}, the assumption on $\exists V_i^*\in C^1$ is adopted mainly to simplify the presentation. We use \(C^1\) assumption for consistency with commonly used formulations in HJB-based  methods~\cite{bandyopadhyay2025lagrangian,cohen2023safe,lewisoptimal}. Theorem 1 can be easily relaxed to the case where \(V_i^*\) is locally Lipschitz, while the controller structure
and the learning algorithm remain unchanged.  The corresponding technical
statement is provided in Remark~\ref{rem3} and Corollary~\ref{the2} of
Appendix\ref{App-proof-the}. Moreover, in implementation, the desired properties of the value
function and the graph CBF are promoted by the structured losses introduced in
Section~\ref{Section4}. Further details on the numerical approximation are
given in Section~\ref{Section4} and Remark~\ref{rem2} of
Appendix\ref{App-proof-the}.
\end{remark}

To implement the analytical controller \eqref{us}, we need to address the following three practical challenges: (i) the Lagrange multiplier $\lambda_i^*$ used in controller \eqref{us} involves neighbor control coupling, hindering distributed execution; (ii) the value function $V_i^*$ is difficult to be obtained; and (iii) a valid graph CBF $h$ is unknown for unknown environments, unlike known safety constraints used in \cite{bandyopadhyay2025lagrangian,cohen2023safe, almubarak2021hjb}.

\section{HJB-GNN: Physics-Informed Safe Learning with GNNs}\label{Section4}

This section develops a novel HJB-GNN learning framework to approximate the safe optimal solution \eqref{us}.

\subsection{GNN-based Approximation}\label{Section4.1}

GNNs serve as a powerful tool for developing learning-based methods  over graph-structured data in MAS. We leverage GNNs to parameterize both a graph CBF $h_{\vartheta}$ and a distributed safe policy $\pi_\varphi$ with the network parameters $\vartheta$ and $\varphi$. 
Unlike conventional fixed-input neural network architectures \cite{zhang2025towardtro,saravanos2023distributed,Vinod2024tcst,yan2025tnnls}, GNNs inherently support variable-sized neighbor sets $\mathcal{N}_i$ and ensure generalizability to arbitrary swarm scales without requiring padding or truncation \cite{Ma2023ITSC,yu2023learning}. 
We parameterize a value function $V_\theta(\bold{e}_i)$ using a MLP with a network parameter $\theta$, since it only depends on goal-reaching error $\bold{e}_i$ not graph topology.

For $h_\vartheta$ and $\pi_\varphi$, we employ a graph attention-based GNN to encode local interactions within the topology graph $G$, augmented with a goal node for each agent and an edge connecting the agent to its goal. The input feature for agent $i$ is denoted as $\bold{\eta}_i=(\bold{\eta}_{i1},...,\bold{\eta}_{i|\mathcal{N}_{i}|})$, where for each $j\in \mathcal{N}_{i}$, the feature $\bold{\eta}_{ij}$ concatenates the node features to encode node type (agent, goal, or obstacle) and the edge feature containing the relative position $\bold{p}_j-\bold{p}_i$. This architecture ensures that the policy is \textit{distributed}, as the computation of $\pi_\varphi$ for agent $i$ only relies on its local neighborhood $\mathcal{N}_i$. Detailed GNN architecture is illustrated in Appendix\ref{App-GNN}.

\subsection{Training Process}\label{Section4.2}
In this section, we design three loss functions to train three networks $V_\theta(\bold{e}_i)$, $h_{\vartheta}(\bold{\eta}_i)$, and $\bold{\pi}_{\varphi}(\bold{\eta}_i)$, respectively. 
We present the detailed design of each loss function below. For convenience, let $L_{*}=\sum_{i\in V_a}L_{*i}$ with $*\in\{v,h,u\}.$

\textbf{Value Function Loss.} We design the value function loss $L_{vi}(\theta,\varphi)$ consisting of \textit{candidate Lyapunov function} loss $L_{cvi}(\theta)$ and squared \textit{Bellman error} loss $L_{hvi}(\theta,\varphi)$, i.e.,
\begin{align}\label{lvi}
L_{vi}(\theta,\varphi)=b_{v1}L_{cvi}(\theta)+b_{v2}L_{hvi}(\theta,\varphi), 
\end{align}
where $b_{v1},\ b_{v2}\in\mathbb{R}_{>0}$ are two constant weights. Minimizing $L_{cvi}(\theta)$ encourages $V_{\theta}(\bold{e}_i)$ to act as a candidate Lyapunov function of the system of goal-reaching error. According to the requirement that a candidate Lyapunov function be both positive definite and radially unbounded in~\cite{khalil1996nonlinear}, we design $L_{cvi}(\theta)$ to enforce these properties:
\begin{align}\label{Lvir}
L_{cvi}(\theta)
=~&\max\big\{0,\alpha_1(\|\bold{e}_i\|)-V_{\theta}(\bold{e}_i)\big\}\nonumber\\
&+\max\big\{0,V_{\theta}(\bold{e}_i)-\alpha_2(\|\bold{e}_i\|)\big\},
\end{align}
where $\alpha_1$, $\alpha_2$ are of class $\mathcal{K}_\infty$ functions satisfying $\alpha_1(s)<\alpha_2(s)$ for any $s$.

Based on (\ref{optimal-H}) and (\ref{optima-lag}), the Bellman error $\delta_i(\theta,\varphi)$ has 
\begin{align}\label{bell}
&\delta_i(\theta,\varphi)\nonumber\\
=~&H_i(\bold{e}_i,\bold{\pi}_{\varphi}(\bold{\eta}_i),\nabla_{\bold{e}_i} V_\theta(\bold{e}_i))
-H_i(\bold{e}_i,\bold{u}_{i}^*,\nabla_{\bold{e}_i} V_i^*(\bold{e}_i))
\nonumber\\
=~&\nabla_{\bold{e}_i} V_\theta(\bold{e}_i)\big(\bold{f}(\bold{x}_i)+\bold{g}(\bold{x}_i)\bold{\pi}_{\varphi}(\bold{\eta}_i)\big)+r_{i}(\bold{e}_i,\bold{\pi}_{\varphi}(\bold{\eta}_i)).
\end{align}
Minimizing the loss $L_{hvi}(\theta,\varphi)=\delta_i^2(\theta,\varphi)$ in (\ref{lvi}) encourages the satisfaction of HJB optimality condition in an approximate sense, i.e., $H_i(\bold{e}_i,\pi_{\varphi}(\bold{\eta}_i),\nabla_{\bold{e}_i} V_\theta(\bold{e}_i))=0$, guiding $V_\theta(\bold{e}_i)$ toward the true value function $V_i^*(\bold{e}_i)$. Such design also encourages the time derivative of function $V_\theta(\bold{e}_i)$, i.e., $\nabla_{\bold{e}_i} V_\theta(\bold{e}_i)(\bold{f}(\bold{x}_i)+\bold{g}(\bold{x}_i)\pi_{\varphi}(\bold{\eta}_i))$, to be negative definite along the system trajectory of goal-reaching error. 
Together with $L_{cvi}(\theta)$, it aligns with the fact that the value function $V_{i}^*(\bold{e}_i)$ serves as a Lyapunov function for the goal-reaching error system in Theorem~\ref{the1}, and $V_\theta(\bold{e}_i)$ is trained to inherit such property. This  Lyapunov function property promotes that each agent can reach to its goal position.

\textbf{CBF Loss.} We construct two datasets $D_i^C$ and $D_i^A$ consisting of input features labeled as safe and unsafe, respectively, as detailed in Appendix\ref{App-GNN-loss} and following the procedure in \cite{zhang2025tro}. The graph CBF loss $L_{hi}(\vartheta,\varphi)$ encourages the satisfaction of safety constraint (\ref{CBF_N_R}) and the safety requirement from $\mathcal{H}_{N,i} \subseteq \mathcal{S}_{i}$ in Theorem~\ref{the1}, i.e., 
\begin{align}\label{lhi}
&L_{hi}(\vartheta,\varphi)\nonumber\\
=&\sum_{\bold{\eta}_i\in D_{i}^C}\max\big\{0,\varsigma-h_{\vartheta}(\bold{\eta}_i)\big\}+\sum_{\bold{\eta}_i\in D_{i}^A}\max\big\{0,\varsigma+h_{\vartheta}(\bold{\eta}_i)\big\}\nonumber\\
&+b_h\sum_{\bold{\eta}_i}\max\big\{0,\varsigma-\dot{h}_{\vartheta}(\bold{\eta}_i)-\alpha(h_{\vartheta}(\bold{\eta}_i))\big\},
\end{align}
where $\varsigma\in\mathbb{R}_{>0}$ is a constant weight to  encourage these conditions to strictly be satisfied, and $b_h\in\mathbb{R}_{>0}$ is a constant weight. Minimizing the loss \eqref{lhi} encourages that each agent maintains safety distances with other agents and obstacles.

\textbf{Controller Loss.} We design the controller loss function $L_{u i}(\theta,\vartheta,\varphi)$ as: 
\begin{align}\label{lci}
L_{u i}(\theta,\vartheta,\varphi)=
L_{\pi i}(\theta,\vartheta,\varphi)+L_{hi}(\vartheta,\varphi)+L_{hvi}(\theta,\varphi).
\end{align}
Here, minimizing the CBF loss $L_{hi}(\vartheta,\varphi)$ promotes the controller $\pi_{\varphi}(\eta_i)$ to satisfy the safety constraints. 
Minimizing $L_{hvi}(\theta,\varphi)$ enforces 
the satisfaction of HJB optimality condition and guides the time derivative of value function to be negative definite and the controller to be stabilizing.

We introduce the loss $L_{\pi i}(\theta,\vartheta,\varphi)$ to minimize the deviation between $\pi_\varphi(\bold{\eta}_i)$ and the approximated controller $\bold{\hat{u}}_{i}$ of safe optimal controller \eqref{us}, without requiring the QP-CBF controller in \cite{zhang2025tro}:
\begin{align}
L_{\pi i}(\theta,\vartheta,\varphi)=b_{\pi}\|\bold{\pi}_{\varphi}(\bold{\eta}_i)-\bold{\hat{u}}_{i}\|,\label{lui}
\end{align}
where $b_{\pi}\in\mathbb{R}_{>0}$ a constant weight, the approximated controller $\bold{\hat{u}}_{i}$ of \eqref{us} is 
\begin{align}\label{appro-us} 
 \bold{\hat{u}}_{i}=-\frac{1}{2}\bold{R}^{-1}\bold{g}^T(\bold{x}_i)\big(\underbrace{\nabla_{\bold{e}_i}  V_\theta(\bold{e}_i)}_{\text{Goal-reaching}}-\hat{\lambda}_i\underbrace{\nabla_{\bold{\eta}_i} h_{\vartheta}(\bold{\eta}_i)}_{\text{Safety}}\big),  
\end{align}
and the approximated Lagrange multiplier $\hat{\lambda}_i$ is defined in \eqref{lagrange1} when replacing $V_i^*(\bold{e}_i)$ and $h(\bold{\bar{x}}_{\hat{\mathcal{N}}_i})$ with $V_\theta(\bold{e}_i)$ and $h_{\vartheta}(\bold{\eta}_i)$, respectively. 
By minimizing the deviation between $\pi_\varphi(\bold{\eta}_i)$ and $\bold{\hat{u}}_{i}$, the loss (\ref{lui}) guides $\bold{\pi}_{\varphi}(\bold{\eta}_i)$ to inherit the safety, goal-reaching performance, and optimality structure encoded in the KKT-based controller. This design achieves the adaptive trade-off between safety and goal-reaching performance via $\hat{\lambda}_i$, and distributed control $\pi_\varphi(\bold{\eta}_i)$ instead of centralized control \eqref{us}.

\begin{algorithm}[htbp]
\caption{HJB-GNN Learning Algorithm for Multi-Agent Navigation.}
\begin{algorithmic}[1]
\State \textbf{Initialization:} 
Network parameters $\theta$, $\vartheta$, $\varphi$, the compact set $\mathcal{X}$, and the number of samples $n_s$
\For{each iteration} 
\State Collect $D_i^A$ and $D_i^C$ using policy $\pi_{\varphi}(\eta_i)$, $\forall i\in V_a$
\State Collect $D_v$ via uniformly sampling $n_s$ data from $\mathcal{X}$
\Statex\emph{\# Phase 1: update $V_{\theta}(\bold{e}_i)$ and $h_{\vartheta}(\eta_i)$ under fixed $\pi_{\varphi}(\eta_i)$}
\Repeat 
\State Update $\theta$ by minimizing $L_v(\theta,\varphi)$
\State Update $\vartheta$ by minimizing $L_h(\vartheta,\varphi)$
\Until $\mathbb{E}_{\bold{e}_i\sim D_v}\delta_i^2(\theta,\varphi)$ is reduced for all $ i\in V_a$
\Statex\emph{\# Phase 2: update $\pi_{\varphi}(\eta_i)$ and $h_{\vartheta}(\eta_i)$ under fixed $V_{\theta}(\bold{e}_i)$}
\Repeat 
\State Update $\varphi$ by minimizing $L_u(\theta,\vartheta,\varphi)$
\State Update $\vartheta$ by minimizing $L_h(\vartheta,\varphi)$
\Until 
$\mathbb{E}_{\bold{e}_i\sim D_v}\dot{V}_\theta(\bold{e}_i(t))<0$ for all $i\in V_a$
 \EndFor
\end{algorithmic}
\label{alg1}
\end{algorithm}

\textbf{HJB-GNN Algorithm.} The overall algorithm is given in Algorithm~\ref{alg1}. We perform the training by minimizing the corresponding loss functions via gradient descent, during which $\theta$, $\vartheta$, and $\varphi$ are updated correspondingly. Specifically, we can randomly initialize $\theta$, $\vartheta$ and conduct a warm-up phase by iteratively minimizing the value function loss~(\ref{lvi}) to yield an admissible initial controller $\pi_\varphi$. Then, we adopt two phases in each iteration: 
\textbf{(i)  Phase~1:} With the controller $\pi_{\varphi}(\eta_i)$ fixed, the value function and graph CBF are updated to evaluate the current controller by minimizing the loss $L_{v}(\theta,\varphi)$ and $L_h(\theta,\varphi)$, respectively. 
Phase~1 terminates when the empirical expectation of the squared Bellman error $\mathbb{E}_{\bold{e}_i\sim D_v}\delta_i^2(\theta,\varphi)$ for all $i\in V_a$ is reduced. 
\textbf{(ii) Phase 2:} With the value function fixed, the controller is improved by minimizing the controller loss, while the graph CBF is further refined. This phase encourages the controller to be stabilizing and safe with respect to the current value function. Phase~2 terminates when the empirical expectation of the Hamiltonian becomes non-positive, i.e., 
$\mathbb{E}_{\bold{e}_i\sim D_v}\dot{V}_{\theta}(\bold{e}_i(t))<0$ 
for all $ i\in V_a$, indicating the controller to be stabilizing. By alternating between these two phases, the algorithm progressively yields a distributed safe policy $\pi_{\varphi}(\eta_i)$ for multi-agent navigation.

 \begin{figure}[htpb]
  \centering
\includegraphics[width=0.45\textwidth, trim=0cm 0 0 0, clip]{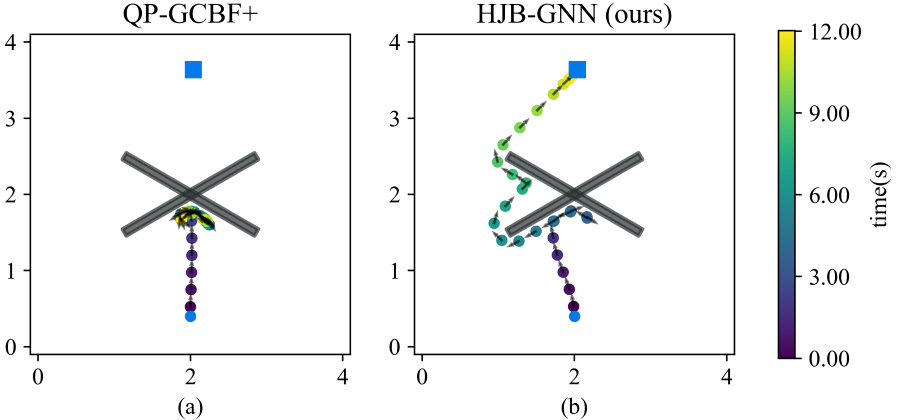}
 \caption{Comparison between QP-GCBF+ of \cite{zhang2025tro} in (a) and our HJB-GNN in (b). An agent (a gradient-colored circle) aims to reach the goal (blue square) while avoiding static obstacles (gray rectangles) within a $4{\rm m}\times4{\rm m}$ region. Black arrow indicates the velocity direction of agent. The QP-GCBF+ has deadlock near obstacles, and the HJB-GNN achieves goal-reaching without collision. }
\label{fig.1}
 \end{figure}

\emph{Reduced Deadlocks.} 
Our HJB-GNN algorithm can reduce deadlocks 
that can be caused by the use of short-horizon QP solvers  \cite{Mrdjan2024cst,Mestres2024ral,Wang2017tro,zhang2025tro}.  
We present a comparison between state-of-the-art QP-GCBF+ method in \cite{zhang2025tro} and our HJB-GNN approach in Fig.~\ref{fig.1}. The QP-GCBF+ method in \cite{zhang2025tro} suffers from deadlocks near obstacles, our HJB-GNN controller ensures a collision-free trajectory to the goal. 

\section{Simulations and Hardware Experiments}\label{Section5}
In this section, we design extensive simulations and hardware experiments to answer the three main questions:

\begin{figure}[htpb]
\vspace{-2mm}
 \centering
\includegraphics[scale=1.95]{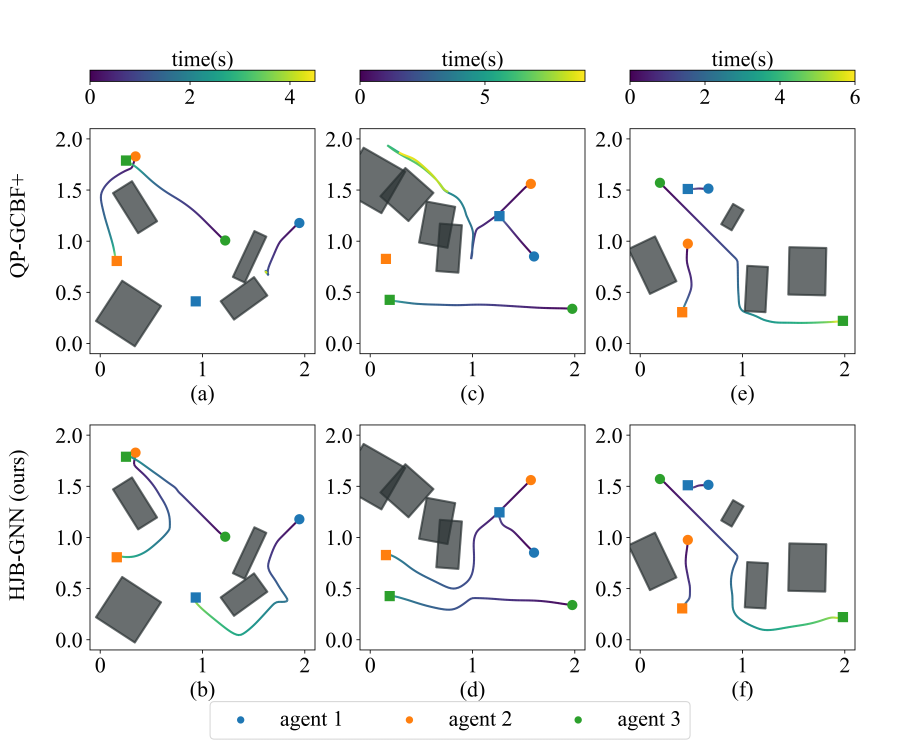}
 \caption{Comparisons between the QP-GCBF+ method ((a), (c), and (e)) in \cite{zhang2025tro} and our HJB-GNN approach ((b), (d), and (f)) under three cases with four agents (circles indicating their initial positions) and four static obstacles (gray rectangles) in a $2{\rm m}\times2{\rm m}$ region. Agent 1 in (a) and agent 2 in (c) experience deadlock, and agent 3 in (e) collides with an obstacle. All agents safely reach their respective goals (squares with the same colors as agents) in (b), (d), and (f). }\label{fig.7}
 \end{figure}

\textbf{(Q1)} Does the learned policy from HJB-GNN algorithm reduce deadlocks and collisions in existing short-horizon QP-CBF methods with predefined nominal controllers?

\textbf{(Q2)} How well does the learned policy from HJB-GNN algorithm scale to unseen dense environments and generalize to much larger MAS?

\textbf{(Q3)} Can HJB-GNN algorithm is robust enough for effective deployment on physical robotic platforms?

Details on the simulations and hardware implementation, hyperparameters, and more evaluation results are given in Appendixes\ref{App-simulation-setup} and \ref{App-hardware}.

\subsection{Simulations Experiments}\label{Section5.1}
We compare our HJB-GNN approach and state-of-the-art QP-GCBF+ method in \cite{zhang2025tro} on both 2D linear double integrator and 3D nonlinear Crazyflie drone systems as described in \cite{zhang2025tro}. Two metrics are used to evaluate performance: (i) \textbf{safety rate}, defined as the proportion of agents that avoid all collisions throughout the simulation; and (ii) \textbf{safe-reaching rate}, defined as the proportion of agents that both remain collision-free and successfully reach their goals. For each simulation, the performance is assessed by randomly choosing 32 initial positions of agents and goal positions, and safety rates and safe-reaching rates are reported as the mean and standard deviation over these 32 instances. All experiments are trained with 8 agents and 8 static obstacles.

\begin{figure}[htpb]
 \centering
\includegraphics[scale=2.10]{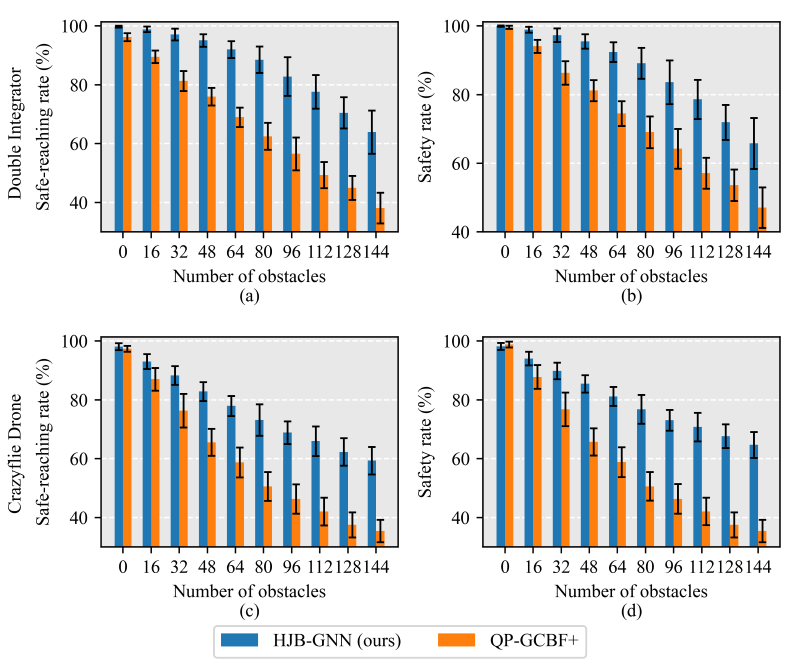}
 \caption{Safe-reaching rates and safety rates for increasing numbers of obstacles and 256 agents under fixed area width. (a)-(b): 2D double integrator, and (c)-(d): 3D Crazyflie drone. }\label{fig.5}
 \end{figure}

   \begin{figure}[htpb]
 \centering
\includegraphics[scale=2.10]{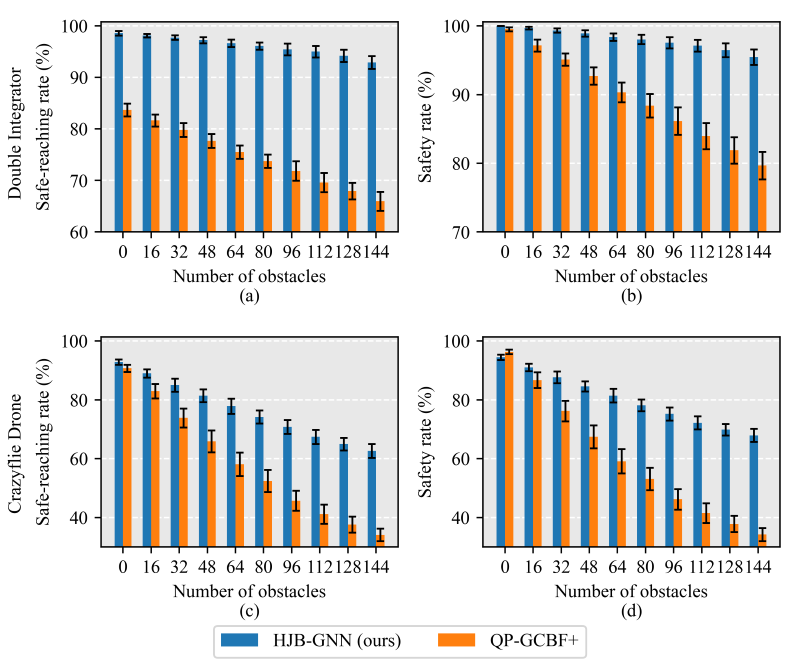}
 \caption{Safe-reaching rates and safety rates for increasing numbers of obstacles and 1024 agents under fixed area width. (a)-(b): 2D double integrator, and (c)-(d): 3D Crazyflie drone. }\label{fig.6}
 \vspace{-3mm}
 \end{figure}

\textbf{(Q1): HJB-GNN approach reduces deadlocks and collisions in the baseline QP-GCBF+ \cite{zhang2025tro}.} 
We conduct three comparative experiments in dense, cluttered environments, shown in Fig.~\ref{fig.7}. The QP-GCBF+ method suffers from deadlocks (see Agent 1 in (a) and Agent 2 in (c) of Fig.~\ref{fig.7}) or collisions (Agent 3 in (e) of Fig.~\ref{fig.7}). In contrast, our HJB-GNN approach ensures that all agents safely reach their goals, shown in Fig.~\ref{fig.7} (b), (d), and (f). In the baseline, the failures stem from the structural limitations caused by short-horizon optimization and non-safety-aware fixed nominal controllers, which restrict the coordination between safety and goal-reaching, thereby easily causing deadlocks and collisions in dense, cluttered environments. In the HJB-GNN, 
we leverage an infinite-horizon optimal formulation and the adaptive trade-off mechanism, which ensure that the learned policy provides more forward-looking and flexible decision-making, thereby reducing deadlocks and collisions.

\textbf{(Q2): The learned policy can scale to unseen dense environments and generalize to much larger MAS.} 
We conduct 
two sets of simulations by fixing number of agents while increasing obstacle density, as illustrated in Fig.~\ref{fig.5} (256 agents) and Fig.~\ref{fig.6} (1024 agents). The results demonstrate that HJB-GNN consistently outperforms the QP-GCBF+ baseline in both safety and safe-reaching rates. 
Notably, the performance advantage becomes more pronounced as obstacle density intensifies, primarily due to the rapid degradation of the baseline method. This failure in the baseline stems from its reliance on precomputed nominal controllers, which restrict the flexibility to adaptively reconcile two competing objectives. Furthermore, its short-horizon optimization often induces deadlocks and overly-conservative behaviors in high-density scenarios.
In contrast, our approach effectively overcomes these limitations by integrating infinite-horizon optimal navigation with a principled adaptive trade-off mechanism. The HJB-GNN approach significantly enhances the policy's scalability and its ability to generalize to large-scale MAS in previously unseen, high-density environments.

\begin{figure*}[htpb]
   \centering
\includegraphics[scale=0.62]{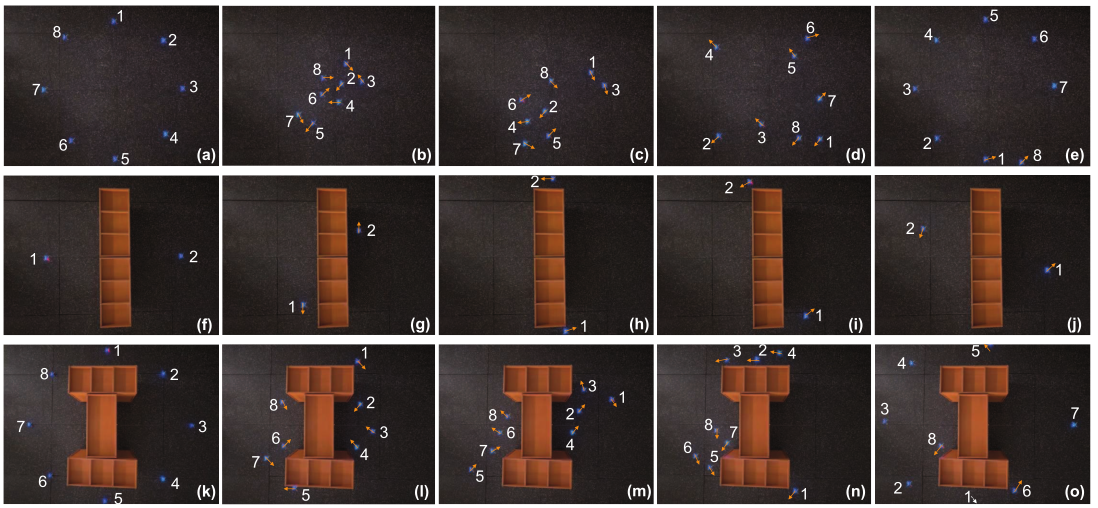}
 \caption{\textbf{Hardware experiments on three antipodal position-swapping tasks.} From left to right: key frames of trajectories across three experiments. Orange arrows denote velocity directions. (a)-(e): Eight Crazyflie drones in an obstacle-free environment. (f)-(j): Two drones navigating around a long static obstacle. (k)-(o): Eight drones navigating among large non-convex obstacles. white dashed arrow in (o) indicates that Drone 1 has reached its goal. }\label{fig.8}
 \vspace{-5mm}
 \end{figure*}

\subsection{Hardware Experiments}\label{Section6}
We demonstrate the applicability and robustness of HJB-GNN by four sets of challenging antipodal position-swapping tasks using a swarm of Crazyflie 2.1 platforms \footnote{\href{https://www.bitcraze.io/products/old-products/crazyflie-2-1/}{https://www.bitcraze.io/products/old-products/crazyflie-2-1/}}, shown in Figs.~\ref{fig-mas} and \ref{fig.8}.

The control architecture for the Crazyflie swarm used in this paper follows an offboard-onboard split, considering the limited computation resources of Crazyflies.
The distributed safe policy is running offboard using a PC that is connected to the Crazyflies using Crazyradio PA. \footnote{\href{https://www.bitcraze.io/products/crazyradio-pa/}{https://www.bitcraze.io/products/crazyradio-pa/}}. 
We use the Crazyswarm package \footnote{\href{https://crazyswarm.readthedocs.io/en/latest/}{https://crazyswarm.readthedocs.io/en/latest/}}
to communicate with Crazyflies based on ROS1, which allows for sending full state control commands to the Crazyflies. Moreover, the learned policy requires only $\sim$~1.5 MB memory per agent and a single forward pass at inference, with no online optimization. The computational cost is minimal; for typical problem sizes, inference can be executed in a few milliseconds on embedded platforms such as NVIDIA Jetson Orin Nano. 

\textbf{(Q3): HJB-GNN algorithm can be effectively deployed on physical Crazyflie drone platforms.} All drones are uniformly initialized on a circle of radius $1{\rm m}$ in (a) of Fig.~\ref{fig-mas} and (a), (f) of Fig.~\ref{fig.8}, and $1.4{\rm m}$ in (k) of Fig.~\ref{fig.8}. \emph{(i) Dense agent environment:} Fig.~\ref{fig.8} (a) contains typical inter-agent interactions and serves as a representative
way to evaluate both safety and goal-reaching capabilities. All drones successfully reach their respective antipodal goal positions without collisions throughout the process, see Fig.~\ref{fig.8} (b)-(e). \emph{(ii) Static large, non-convex obstacles:} Fig.~\ref{fig.8} (f) evaluates the HJB-GNN's capability to avoid deadlocks and ensure safety in the presence of long obstacle. Fig.~\ref{fig.8} (k) increases the difficulty due to the risk of getting trapped or experiencing deadlocks near concave corners. From Fig.~\ref{fig.8} (g)-(j) and Fig.~\ref{fig.8} (l)-(o), all drones safely reach their respective goal positions without collisions and deadlocks, highlighting the robustness of our HJB-GNN approach in cluttered environments with complex static obstacle geometries. \emph{(iii) Large dynamic obstacle:} 
The experiment in Fig.~\ref{fig-mas} (a) contains both a static obstacle and a large dynamic obstacle. 
Our control policy is trained in environments with only static obstacles, since we consider that obstacle velocities are unavailable due to limited sensing capability and thus didn't utilize them. To evaluate the scalability of our HJB-GNN approach, we introduce a dynamic obstacle whose motion is unknown to the agents and whose size is significantly larger than that of a drone. By Fig.~\ref{fig-mas} (b)-(e), all drones successfully avoid obstacles and reach their goal positions, indicating the scalability of HJB-GNN approach to previously unseen dynamic obstacle environments.

\section{Conclusions}\label{Section7}
This paper has proposed HJB-GNN, a novel constrained infinite-horizon optimization-guided learning framework for safe multi-agent navigation in unknown, cluttered, and dense environments. By utilizing the KKT optimality conditions, the proposed framework has solved constrained HJB optimal control problem with graph CBF constraints, and derived 
graph-dependent Lagrange multipliers to achieve adaptive trade-off between safety and goal-reaching behaviors, depending on changing interaction topology graph. The resulting HJB-GNN algorithm 
has jointly learned a value function, a graph CBF, and a distributed policy in an end-to-end manner, without relying on short-horizon QP solvers, predefined nominal controllers, or heuristic safety-task trade-off designs. Such algorithm provided a centralized training and distributed execution manner. Extensive simulation results demonstrated that HJB-GNN approach outperforms the state-of-the-art baseline in terms of safety, scalability, and generalizability, particularly in densely cluttered environments. Moreover, real-world experiments with Crazyflie drone swarms on challenging antipodal position-swapping tasks validated the practical applicability of the HJB-GNN approach.


\section*{Acknowledgment}
This work was supported by the Singapore Ministry of Education Tier 2 Academic Research Fund (T2EP20123-0037, T2EP20224-0035).


\bibliographystyle{plainnat}
\bibliography{references}


\clearpage
 \appendices

\section*{Appendix}



%

\subsection{Related Work} \label{App-related-work} %

\textbf{Handcrafted Safety Constraints}. 
Recent studies enforce safety using handcrafted CBFs embedded in QPs \cite{wu2025smc,Dimitratrust,Dimarogonasa,yan2025tnnls,Mrdjan2024cst,Mestres2024ral,Wang2017tro}, or handcrafted BFs incorporated into HJB-based optimal control formulations \cite{bandyopadhyay2025lagrangian,cohen2023safe, almubarak2021hjb}. Existing HJB-based approaches primarily focus on single-agent systems, and extending them to MASs remains challenging due to the complexity of coupled interactions and safety constraints. Moreover, handcrafted CBFs and BFs are highly sensitive to the chosen analytical forms and tuning parameters. Poorly designed them may lead to overly-conservative behaviors or deadlocks in dense multi-agent environments. In addition, these methods typically rely on explicit geometric knowledge of obstacles, which limits their applicability in unknown environments.

\textbf{Learning-based Multi-agent Navigation}. 
Recent years have witnessed growing interest in learning-based safe control methods for MASs. Several works integrated model predictive control (MPC) into reinforcement learning frameworks to enable safety-aware planning \cite{Vinod2024tcst,Safaoui2024tro,zhang2025towardtro}. Their high computational cost limits scalability in large-scale MASs. Moreover,~\cite{zhang2025towardtro} assumed a fixed number of neighbors per agent, which restricts applicability in dynamic environments with changing interaction topologies.
To improve scalability under changing interaction graphs, recent studies have employed GNNs to parameterize multi-agent navigation policies \cite{Ma2023ITSC,yu2023learning}. However, in these approaches safety is typically promoted only implicitly through loss design, without formal guarantees. To provide explicit safety guarantees, neural CBF methods have been proposed for LiDAR-based safe navigation in unknown environments \cite{DawsonLearning,gaoprovably,zhang2023neural,zhang2025tro,zhang2025discrete}. For example, an online exploration-based control Lyapunov BF controller was developed to adapt to previously unseen environments \cite{gaoprovably}, and a hybrid goal-seeking/exploratory controller was proposed to alleviate deadlocks and improve task success rates \cite{DawsonLearning}.
Despite these advances, most existing neural CBF approaches rely on short-horizon optimization \cite{DawsonLearning,gaoprovably,zhang2025tro}, predefined nominal controllers \cite{zhang2023neural,zhang2025tro}, or manually tuned parameters to balance safety and task performance \cite{zhang2023neural}. As a result, they can suffer from myopic decision-making and degraded safety and task performance in cluttered, dense multi-agent scenarios.

\subsection{Proof of the Theorem~\ref{the1}}\label{App-proof-the}
\emph{Proof:} We first prove the safety of the MAS \eqref{mas}, then analyze the goal-reaching error below. To illustrate the safety, we illustrate the satisfaction of graph CBF constraint \eqref{kkt3}. Noting $\|\nabla_{\bold{x}_i}h(\bold{\bar{x}}_{\hat{\mathcal{N}}_i})\bold{g}(\bold{x}_i)\|\ge \varrho$ for all $\bold{\bar{x}}_{\hat{\mathcal{N}}_i}\in \mathcal{X}^M$ implies  $\omega_{i}>0$. If $\Lambda_{i}\ge0$, then 
\begin{align*}
\dot{h}(\bold{\bar{x}}_{\hat{\mathcal{N}}_i}(t))+\alpha(h(\bold{\bar{x}}_{\hat{\mathcal{N}}_i}(t))=\Lambda_{i}+\lambda^*_i\omega_{i}\ge0,
\end{align*}
by means of \eqref{lagrange1} and $\lambda^*_i=0$ given in (\ref{lag}). 
If $\Lambda_{i}<0$, then 
\begin{align*}
\dot{h}(\bold{\bar{x}}_{\hat{\mathcal{N}}_i}(t))+\alpha(h(\bold{\bar{x}}_{\hat{\mathcal{N}}_i}(t))=\Lambda_{i}+\lambda^*_i\omega_{i}=0,
\end{align*}
by using \eqref{lagrange1} and $\lambda^*_i=-\Lambda_{i}/\omega_{i}$ given in (\ref{lag}). 
In both cases, the graph CBF constraint \eqref{kkt3} is satisfied for any $t\ge t_0$. The function $h(\bold{\bar{x}}_{\hat{\mathcal{N}}_i}(t)$ is $C^1$, applying the comparison lemma~\cite{khalil1996nonlinear} to the constraint \eqref{kkt3} produces that 
\begin{align}\label{t1.2}
h(\bold{\bar{x}}_{\hat{\mathcal{N}}_i}(t_0))\ge0\Rightarrow   h(\bold{\bar{x}}_{\hat{\mathcal{N}}_i}(t))\ge0, \ \forall t\ge t_0,
\end{align}
for any $i\in V_a$. Recall that $\mathcal{H}_{N,i}\subseteq \mathcal{S}_{i}$ for all $i\in V_a$ implies $\mathcal{H}_{N}\subseteq \mathcal{S}$. 
For any $\bold{\bar{x}}(t_0)\in \mathcal{H}_{N}\subseteq \mathcal{S}$, i.e., $h(\bold{\bar{x}}_{\hat{\mathcal{N}}_i}(t_0)\ge0$ for all $i\in V_a$, it follows from (\ref{t1.2}) that the state trajectory of the MAS remains in the safe set, i.e., $\bold{\bar{x}}(t)\in \mathcal{H}_{N} \subseteq \mathcal{S}$ for all $t\ge t_0$.

Then, we analyze the goal-reaching error $\bold{e}_i(t)$ of each agent $i$. Using \eqref{optimal-H}, \eqref{optima-lag}, \eqref{us}, and the definition of  $r_i(\bold{e}_i,\bold{u}_i^*)=\bold{e}_i^T\bold{Q}\bold{e}_i+\bold{u}_i^{*T}\bold{R}\bold{u}_i^{*}$, taking the time derivative of $V_{i}^*(\bold{e}_i(t))$ along the goal-reaching error trajectory produces
\begin{align}\label{t1.1-1}
&\frac{{\rm d}  V_i^*(\bold{e}_i(t))}{{\rm d} t}\nonumber\\
=&
\nabla_{\bold{e}_i(t)} V_i^*(\bold{e}_i(t))\big(\bold{f}(\bold{x}_i(t))
+\bold{g}(\bold{x}_i(t))\bold{u}_{i}^*(t)\big)\nonumber\\
=&
-r_{i}(\bold{e}_i(t),\bold{u}_{i}^*(t))\nonumber\\
\leq&
-\bold{e}_i^T(t)\bold{Q}\bold{e}_i(t).
\end{align}
Since the functions $V_i^*$, $i\in V_a$ are positive definite and $C^1$, we know from the Lyapunov stability theory \cite{khalil1996nonlinear} that $\lim_{t\to+\infty} \bold{e}_i(t)=0$, i.e., each agent can reach its goal position as $t\to+\infty$. This completes the proof of Theorem~\ref{the1}.
\hfill $\square$

\begin{remark}\label{rem2}
Satisfying the theoretical conditions in Theorem~\ref{the1} is encouraged through network design and loss constructions as in~\cite{almubarak2021hjb,bandyopadhyay2025lagrangian,zhang2023neural,zhang2025tro}. We promote \( \mathcal{H}_{N,i} \subset \mathcal{S}_i \) by the first two terms in loss~Eq. \ref{lhi}, which is theoretically rather a design choice of CBF to encode the safe set. The regularity
condition (i.e., $\|\nabla_{\bold{x}_i}h(\bold{\bar{x}}_{\hat{\mathcal{N}}_i})\bold{g}(\bold{x}_i)\|\ge \varrho$ with $\varrho\in\mathbb{R}_{>0}$ for all $\bold{\bar{x}}_{\hat{\mathcal{N}}_i}\in \mathcal{X}^M$) can be encouraged by a loss with small constant $\varrho>0$, which we omit as it has negligible impact. The positive definiteness of \( V_\theta \) is encouraged by loss~Eq. \ref{lci} together with zero biases of all networks. These design choices guide the learned functions toward satisfying the required properties in practice. 
Designing numerical schemes that rigorously satisfy all theoretical properties is generally challenging (which is the focus of numerical PDEs) and is beyond the scope of this paper.    
\end{remark}

\begin{remark}\label{rem3}
As discussed in Remark~\ref{rem1}, Theorem~\ref{the1} can be easily extended from the
classical \(C^1\) setting to a locally Lipschitz optimal value function. Since a 
locally Lipschitz function is continuous and differentiable almost everywhere
in the state space by Rademacher's theorem (Theorem 3.2 of~\cite{evans2025measure}), the HJB-based Lyapunov analysis can
be carried out along the closed-loop trajectory, provided that \(V_i^*\) is
differentiable at \(\mathbf e_i(t)\) for almost all \(t\), and that the
constrained HJB equation holds at those differentiability times. Under this
trajectory-level condition, the controller derived in
Theorem~\ref{the1} is retained almost everywhere, and no modification of the learning algorithmic is required. 
The following Corollary~\ref{the2} formalizes this relaxation and shows that the same safety
and goal-reaching conclusions as those in Theorem~\ref{the1} remain valid. The
proof follows the same arguments as Theorem~\ref{the1}, with the classical
chain rule applied at the almost all differentiability times of \(V_i^*\).
\end{remark}

\begin{corollary} (Locally Lipschitz Extension of Theorem~\ref{the1}) \label{the2}
Consider an arbitrarily sized nonlinear MAS~\eqref{mas}. 
Suppose that there exists a locally Lipschitz, positive definite optimal value
function $V_i^*(\bold{e}_i):\mathbb R^{n_i}\to \mathbb R_{\ge0}$ 
and a graph CBF a graph CBF $h(\bold{\bar{x}}_{\hat{\mathcal{N}}_i})$ satisfying $\mathcal{H}_{N,i}\subseteq \mathcal{S}_{i}$, $ i\in V_a$ and $\|\nabla_{\bold{x}_i}h(\bold{\bar{x}}_{\hat{\mathcal{N}}_i})\bold{g}(\bold{x}_i)\|\ge \varrho$ with $\varrho\in\mathbb{R}_{>0}$ for all $\bold{\bar{x}}_{\hat{\mathcal{N}}_i}\in \mathcal{X}^M$. Assume $V_i^*$ is differentiable at $\mathbf{e}_i(t)$ and satisfies the
HJB equation \eqref{const-hjb} for almost all $t$ along the closed-loop trajectory. Let the controller be given almost everywhere by~\eqref{us}, with the
graph-dependent Lagrange multipliers in~\eqref{lag}, for all \(i\in V_a\).
Then, for any initial condition \(\bar{\mathbf x}(t_0)\in \mathcal H_N\), the
closed-loop system remains safe, i.e., $\bar{\mathbf x}(t)\in \mathcal S,\ \forall t\ge t_0,$ and the goal-reaching error of each agent asymptotically converges to the zero. \end{corollary}


\subsection{GNN Architecture}\label{App-GNN}
Each feature $\bold{\eta}_{ij}=(\bold{v}_i,\bold{v}_j,\bold{\tilde{e}}_{ij})$ concatenates the node features $\bold{v}_i, \bold{v}_j$ and the edge feature $\bold{\tilde{e}}_{ij}$ for $j\in \mathcal{N}_i$. Node feature $\bold{v}_i\in\mathbb{R}^{\rho_v}$, with $\rho_v = 3$, uses one-hot encoding to indicate whether a node represents an agent, a goal, or an obstacle. Edge feature $\bold{\tilde{e}}_{ij}\in\mathbb{R}^{\rho_e}$, with $\rho_e\in\mathbb{R}_{>0}$, includes the relative position $\bold{p}_{ij}=\bold{p}_j-\bold{p}_i$, is critical for enforcing safety constraints. 
Each input feature $\bold{\eta}_{ij}$ of $h_\vartheta$ is first mapped into a latent space by a MLP $\phi_{\vartheta_1}$ with neural network parameter $\vartheta_1$, yielding $\mu_{ij}=\phi_{\vartheta_1}(\bold{\eta}_{ij})$. Neighbor features for each node are then aggregated using a graph attention mechanism implemented via two additional networks $\phi_{\vartheta_2}$ and $\phi_{\vartheta_3}$, producing:
\begin{align}\label{gnn}
\mu_{i}=\sum_{j\in \mathcal{N}_i}\phi_s(\mu_{ij})\phi_{\vartheta_3}(\mu_{ij}),
\end{align}
where the attention weight $\phi_s(\mu_{ij})={\rm softmax}(\phi_{\vartheta_2}(\mu_{ij}))\in[0,1]$ satisfies $\sum_{j\in \mathcal{N}_i}\phi_s(\mu_{ij})=1$. The attention weight $\phi_s(\mu_{ij})$ encodes the importance of node $j$ to agent $i$. 
Then, a last MLP $\phi_{\vartheta_4}$ is applied to process $\mu_{i}$ from (\ref{gnn}) to generate the graph CBF $h(\bold{\eta}_i)=\phi_{\vartheta_4}(\mu_{i})$ for each agent $i\in V_a$.

\subsection{Design of Datasets $D_i^C$ and $D_i^A$ }\label{App-GNN-loss}
At each learning iteration, for the input feature $\bold{\eta}_i$, the policy from the previous iteration is used to roll out the system trajectories for a finite-horizon $T$, and each $\bold{\eta}_i$ is classified according to whether future collisions occur along the system trajectory. Each $\bold{\eta}_i$ that remains collision-free is added to $D_i^C$, those that result in collisions are added to $D_i^A$, and the remaining are left unlabeled.


\subsection{Simulation Setup}\label{App-simulation-setup}
Scalability refers to the ability of the approach to scale to a larger number of agents while maintaining safety and goal-reaching performance. Generalizability refers to the ability of a trained policy to perform well in unseen scenarios that differ from the training distribution, such as different agent densities, goal configurations, and obstacle layouts.

Each agent is simulated to be equipped with a LiDAR sensor for obstacle detection, with observation data collected from evenly spaced LiDAR rays within the sensing radius $R_a$. Each simulation is conducted with $R_a=0.5{\rm m}$ for 2D systems and $R_a=1{\rm m}$ for 3D systems, and a safety distance of $r=0.05{\rm m}$. We perform our training using the Adam optimizer for 1000 steps. A total of 4096 test steps are executed. All training and testing are conducted on a workstation equipped with an AMD Ryzen Threadripper PRO 7965WX CPU and dual NVIDIA RTX 4090 GPUs.

We train $V_\theta(\bold{e}_i)$ using five-layer fully-connected neural networks where each neural network contains four hidden layers with 256 units per layer, activated by rectified linear units. A squared activation function is used for the output layer, and the biases of all networks are set to zero. The training data set $D_v$ is obtained with $n_s=12288$ in Algorithm~\ref{alg1}.  
For the fairness of comparison, we choose the same GNN parameterizations of $h_{\vartheta}(\bold{\eta}_i)$ and $\pi_\varphi(\bold{\eta}_i)$ as those in~\cite{zhang2025tro}. 
Each agent and its goal are initialized by uniformly sampling from $\mathbb{P}_0=[0,l]^{n_d}$, where $l>0$ (area width) is set to $4{\rm m}$ for 2D and $2{\rm m}$ for 3D simulations, respectively. The density of agents is approximately $N/l^{n_d}$, and decreasing $l$ increases the agent density and collision risk. All experiments 
use a time step of $0.03{\rm s}$. We note that the performance of our approach is not sensitive to the choice of hyperparameters, and consistent results are observed across a wide range of parameter values. 

\subsubsection{System Models}\label{app-e-1}We conduct simulations for the proposed HJB-GNN algorithm on 2D nonlinear unmanned surface vessel and comparative simulations on 2D linear double intergrator and 3D nonlinear Crazyflie drone, respectively. To ensure fair comparison, the models used in the comparative simulations follow those in the baseline method~\cite{zhang2025tro}. 
According to \cite{namerikawa2024equivalence}, the dynamic of each unmanned surface vessel agent $i\in V_a$ can be described by
\begin{align}\label{usv}
\bold{\dot{x}}_i=\begin{bmatrix}
\cos(\psi_i) & -\sin(\psi_i) & 0 \\
\sin(\psi_i) & \cos(\psi_i) & 0 \\
0 & 0 & 1 \\
\end{bmatrix}\begin{bmatrix}
v_i \\
w_i \\
r_i \\
\end{bmatrix},
\end{align}
where the state vector $\bold{x}_i=[p_i^x,p_i^y,\psi_i]^T$ includes the positions $p_i^x,\ p_i^y$ and the heading $\psi_i$. The control input is denoted as $\bold{u}_i=[v_i,w_i,r_i]^T$ that contains longitudinal velocity $v_i$, lateral velocity $w_i$, and angular velocity $r_i$. The edge information is defined as $\bold{\tilde{e}}_{ij} = \bold{x}_j - \bold{x}_i$, for all $i,j \in V_a$.

\begin{table}[ht]
    \centering
    \caption{The hyperparameters used in loss functions (\ref{lvi}), (\ref{lhi}), and (\ref{lui}) for all three system models.}
    \label{tab:hyp}
    \begin{tabular}{c|c|c|c|c}
        \hline
        & $b_{v1}$ & $b_{v2}$ & $b_h$ & $b_\pi$ \\
        \hline
        \makecell{Unmanned surface \\ vessel} & $5\times10^{-5}$ & $10^{-5}$ & $10^{-2}$ & $7\times10^{-4}$ \\
        Double integrator        & $0$ & $10^{-3}$ & $10^{-2}$ & $10^{-4}$ \\
        Crazyflie drone          & $10^{-5}$ & $10^{-4}$ & $10^{-2}$ & $10^{-4}$\\
        \hline
    \end{tabular}
\end{table}


\subsubsection{Hyperparameters Details}\label{app-e-2}\label{App1}
In our simulations, we choose $T=32$, $\alpha(s)=s$, and $\varsigma=0.02$ in loss function (\ref{lhi}) for all system models. Other hyperparameters in the loss functions (\ref{lvi}), (\ref{lhi}), and (\ref{lui}) for all three system models are set in Table~\ref{tab:hyp}. In Table~\ref{tab:hyp}, for the double integrator system, the candidate Lyapunov function loss term is omitted, i.e., $b_{v1}=0$, due to its linear dynamic and structural simplicity.
 The training remains stable with the remaining loss terms in (\ref{lvi}). 
The initial learning rates for the policy, value, and CBF networks in all three system models are $10^{-4}$ and decay exponentially with training steps.

\subsubsection{Simulations on Unmanned Surface
Vessel}\label{app-e-3}
We evaluate our HJB-GNN algorithm on 2D nonlinear unmanned surface vessel through three types of simulations: (i) four agents with four obstacles, (ii) obstacle-free case with increasing numbers of agents, and (iii) fixed number of agents case with increasing numbers of obstacles. All simulation results are discussed below.

 \begin{figure*}[htpb]
 \centering
\includegraphics[scale=1.80]{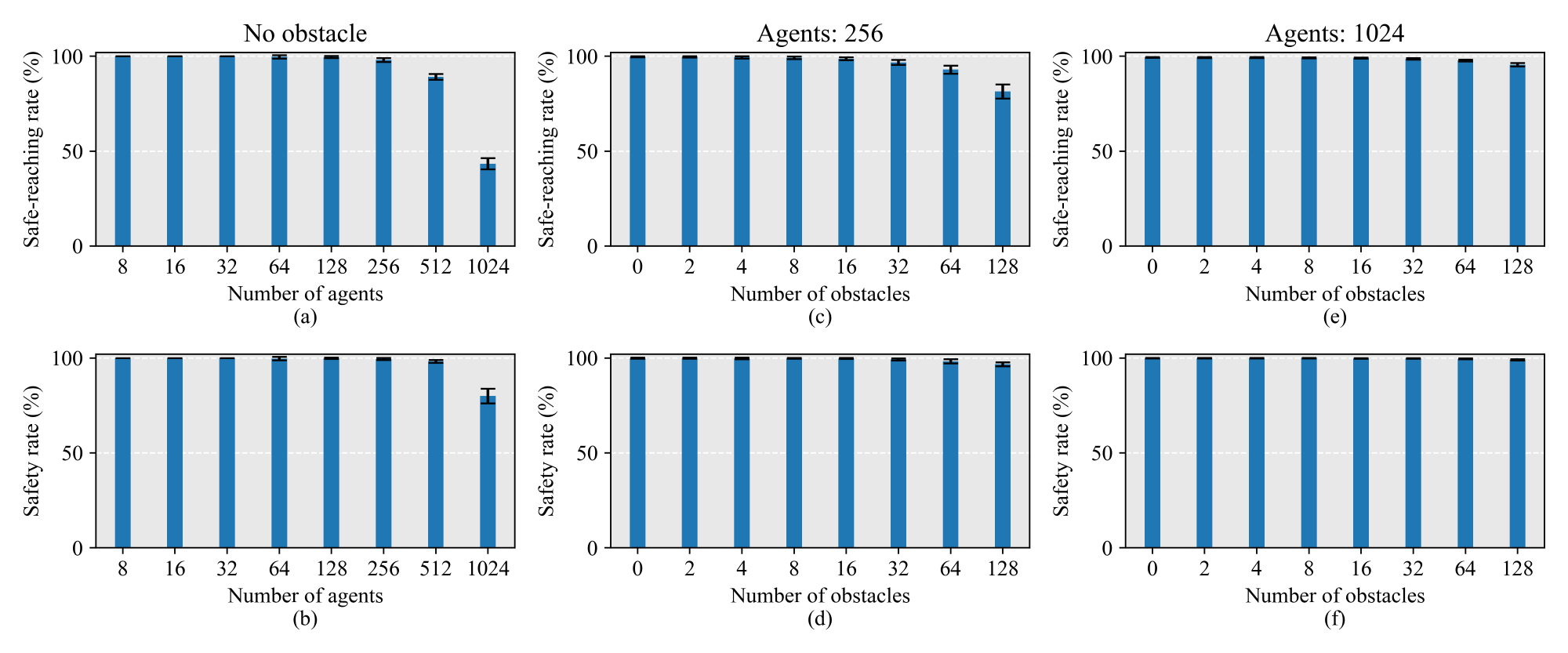}
 \caption{Safe-reaching rates ((a), (c), (e)) and safety rates ((b), (d), (f)) for 2D unmanned surface vessel under fixed area width. (a)-(b): no obstacle with increasing numbers of agents, (c)-(d): 256 agents with increasing numbers of obstacles, and (e)-(f): 1024 agents with increasing numbers of obstacles. 
 }\label{fig.2}
 \end{figure*}

\emph{\textbf{(i) Four Agents with Four Obstacles:}} 
The simulation results are plotted in Fig.~\ref{fig.3} under a $2{\rm m}\times1.5{\rm m}$ environment with four agents and four static obstacles, discussed before Theorem~\ref{the1}.

\emph{\textbf{(ii) Obstacle-Free Cases with Increasing Numbers of Agents:}}
To evaluate scalability and generalizability of our HJB-GNN approach in collision-avoidance among agents, some simulations are conducted in obstacle-free environments with fixed area width $l = 8{\rm m}$ while increasing the number of agents $N$ from 8 to 1024. Here, the agent density is increased more than 100-fold. 
All simulation results are presented in Fig.~\ref{fig.2} (a) and (b), respectively. 
The proposed HJB-GNN approach achieves nearly 
$100\%$ safety and safe-reaching rates when the number of agents is up to 512, demonstrating strong scalability in large-scale MAS. Even when the agent number increases to 1024, leading to significantly higher agent density and more frequent interactions, the approach still maintains a safety rate close to 
$80\%$ and a safe-reaching rate around $50\%$, showcasing its robustness under extreme scalability challenges.

\emph{\textbf{(iii) Fixed Number of Agents with Increasing Numbers of Obstacles:}}
To evaluate the HJB-GNN approach in both obstacle-avoidance and collision-avoidance among agents, we conduct simulations under two configurations: (i) 256 agents and $l = 16{\rm m}$, and (ii) 1024 agents and $l = 32{\rm m}$. Each agent employs 32 evenly spaced LiDAR rays for obstacles detection, and each obstacle is cuboids and its side length is uniformly sampled from $[0.1, 0.6]$m.  
All simulation results are presented in Fig.~\ref{fig.2} (c)-(d) for 256 agents and in Fig.~\ref{fig.2} (e)-(f) for 1024 agents, respectively. It is observed that both the safety rates and safe-reaching rates are consistently near $100\%$ across most scenarios. Even in the high obstacle density scenario involving 256 agents and 128 obstacles, our approach still maintains a safety rate close to $100\%$ and a safe-reaching rate of approximately $82\%$, highlighting its generalizability and scalability under dense obstacle environments.

\subsubsection{Comparative Simulations with \cite{zhang2025tro}}
To evaluate the performance of the proposed HJB-GNN approach in comparison with the baseline QP-GCBF+ method in \cite{zhang2025tro}, three sets of comparative simulations, similar to Appendix\ref{app-e-3}, are conducted on both 2D linear double integrator and 3D nonlinear Crazyflie drone systems.

 \emph{\textbf{(i) Four Agents with Four Obstacles:}}
We conduct three comparative simulations under four agents and four static obstacles in a $2{\rm m}\times2{\rm m}$ region, shown in Fig.~\ref{fig.7} and discussed in Section~\ref{Section5.1}.

  \begin{figure}[htpb]
 \centering
\includegraphics[scale=2.10]{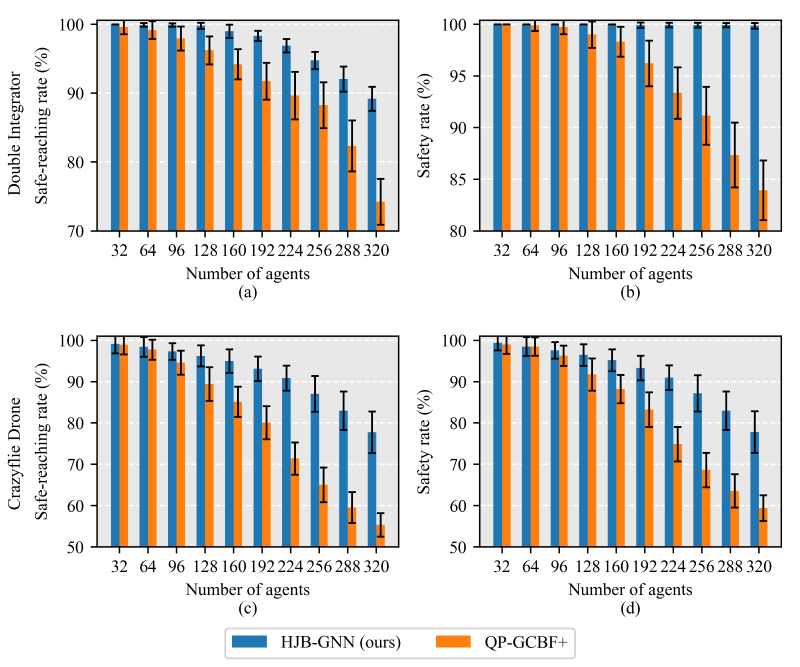}
 \caption{Safe-reaching rates and safety rates for increasing numbers of agents and no obstacle under fixed area width. (a)-(b): 2D double integrator, and (c)-(d): 3D Crazyflie drone. }\label{fig.4}
 \vspace{-4mm}
 \end{figure}

\emph{\textbf{(ii) Obstacle-Free Cases with Increasing Numbers of Agents:}} 
This simulation compares our HJB-GNN approach with QP-GCBF+ method in obstacle-free environments where the number of agents $N$ increases from 32 to 320. The area width is set as $l = 4.5{\rm m}$ for 2D double integrator and $l = 2{\rm m}$ for 3D Crazyflie drone. This setup evaluates collision-avoidance scalability and generalizability by increasing agent density tenfold, where the maximum agent density is larger than that is set in obstacle-free cases of \cite{zhang2025tro}. All simulation results are presented in Fig.~\ref{fig.4} (a)-(b) for double integrator and Fig.~\ref{fig.4} (c)-(d) for Crazyflie drone, respectively. 
It is observed that 
for the linear double integrator, our HJB-GNN approach achieves safety rates close to $100\%$ and safe-reaching rates exceeding $89.3\%$, while QP-GCBF+ yields safety rates below $85\%$ and safe-reaching rates below $75\%$ in the densest agent scenarios. For more challenging nonlinear Crazyflie drone model, our HJB-GNN approach maintains safety and safe-reaching rates above $90\%$ in most cases and above $78\%$ even under the densest agent settings. In contrast, the QP-GCBF+ method suffers significant performance degradation, with safety rates dropping below $60\%$ and safe-reaching rates falling below $55\%$ when the number of agent increases to 320. It is mainly since in contrast with the baseline method, the HJB-GNN removes the structure constraints caused by predefined nominal controllers and enables more effectively coordination between collision-avoidance and goal-reaching control, thereby reducing deadlocks and collisions in dense, cluttered environments.

\emph{\textbf{(iii) Fixed
Number of Agents Cases with Increasing Numbers of Obstacles:}} 
In this simulation, the number of agents $N$ and the area width $l$ remain constant, while the number of obstacles is increased from 0 to 144. Two cases of 256 agents and 1024 agents are tested for 2D double integrator and 3D Crazeflie drone, respectively. For double integrator, the simulations are conducted with $l = 8{\rm m}$ for 256 agents and $l = 16{\rm m}$ for 1024 agents. Each obstacle is a cuboid with side length uniformly sampled from $[0.1, 0.6]$m, and each agent uses the same LiDAR configuration as unmanned surface vessel. For Crazeflie drone, the simulations are conducted with $l = 4{\rm m}$ for 256 agents and $l = 5{\rm m}$ for 1024 agents. Obstacles are modeled as spheres with radii uniformly drawn from $[0.15, 0.3]$m, and each agent employs 130 evenly spaced LiDAR rays for obstacle detection. 
All simulation results are presented under 256 agents and 1024 agents in Figs.~\ref{fig.5} and \ref{fig.6}, respectively, discussed in Section~\ref{Section5.1}.

\subsubsection{Other Comparative Simulations with \cite{zhang2025discrete} and \cite{zhang2025defmarl}}\label{app:additional-baselines}
We provide additional comparisons with two recent learning-based safe multi-agent navigation baselines, DGPPO~\cite{zhang2025discrete} and Def-MARL~\cite{zhang2025defmarl}, in addition to QP-GCBF+~\cite{zhang2025tro}. 
We use the official implementations of DGPPO and Def-MARL and train them for \(15000\) optimization steps, which is substantially longer than the \(1000\) steps used for HJB-GNN, to provide a strong baseline evaluation. 
The evaluation follows Fig.~\ref{fig.4}(a): all methods are trained with 8 agents and tested without retraining from 32 to 320 agents, corresponding to \(1.6\times\) to \(16\times\) higher density than the training setting.


\begin{table}[ht]
\centering
\caption{Safe-reaching rates ($\%,\ \uparrow$) under increasing number of agents from 32 to 320 under double-integrator system.}
\label{tab:app_additional_baselines}
\setlength{\tabcolsep}{2pt}
\begin{tabular}{c|cccccccccc}
\hline
Agents & 32 & 64 & 96 & 128 & 160 & 192 & 224 & 256 & 288 & 320 \\
\hline
\textbf{HJB-GNN}   
& \textbf{99.65} & \textbf{99.51} & \textbf{99.25} & \textbf{99.09} & \textbf{98.29} 
& \textbf{96.68} & \textbf{96.17} & \textbf{94.89} & \textbf{92.49} & \textbf{89.59} \\
QP-GCBF+  
& 99.55 & 98.25 & 97.65 & 96.29 & 95.26 
& 92.58 & 89.26 & 87.39 & 83.49 & 74.45 \\
DGPPO     
& 72.72 & 28.36 & 6.75 & 3.75 & 2.05 
& 1.50 & 1.07 & 0.86 & 0.63 & 0.52 \\
Def-MARL  
& 32.78 & 9.21 & 1.29 & 0.45 & 0.36 
& 0.15 & 0.10 & 0.08 & 0.05 & 0.03 \\
\hline
\end{tabular}
 \vspace{-2mm}
\end{table}

The results in Table~\ref{tab:app_additional_baselines} show that HJB-GNN's advantage is more pronounced as agent
density increases. Moreover, physics-informed methods HJB-GNN and QP-GCBF+ exhibit significantly stronger generalizability than purely data-driven RL methods:  DGPPO and Def-MARL, consistent with the conclusion of \cite{zhang2025discrete}, which states performance degradation due to distribution shift in their data-driven method.

\subsubsection{Ablation Study on Loss Weights}
\label{app:loss-weight-ablation}
We evaluate the sensitivity of HJB-GNN to three key loss weights, \(b_{v2}\), \(b_h\), and \(b_\pi\), by varying each of them over \(\{0.25,0.5,1,2,4\}\) times its default value in Table~\ref{tab:hyp} while keeping the others fixed. 
This gives a \(16\times\) variation range for each tested weight. 
All models are trained with 8 agents and tested without retraining on 256 agents, corresponding to an \(8\times\) higher density than training.

HJB-GNN remains robust over the tested variations, with interpretable effects. 
Increasing \(b_{v2}\) reduces the norm of Bellman error from \(0.81\) to \(0.32\), with a slight improvement in reaching rate from \(92.8\%\) to \(98.6\%\) and a mild decrease in safety rate from \(98.6\%\) to \(92.2\%\). 
Increasing \(b_h\) improves safety rate from \(94.5\%\) to \(98.8\%\), while reducing reaching rate from \(98.8\%\) to \(91.9\%\), reflecting the expected safety-reaching trade-off. 
Varying \(b_\pi\) yields stable safe-reaching rates within \(91.3\%\)-\(98.8\%\). 
These results indicate that the performance is robust to substantial loss-weight variations and that the success of HJB-GNN primarily stems from its
principled constrained HJB-informed structure.

\subsection{Experiments Setup}\label{App-hardware}

\begin{figure}[!t]
    \centering
    \includegraphics[width=\linewidth]{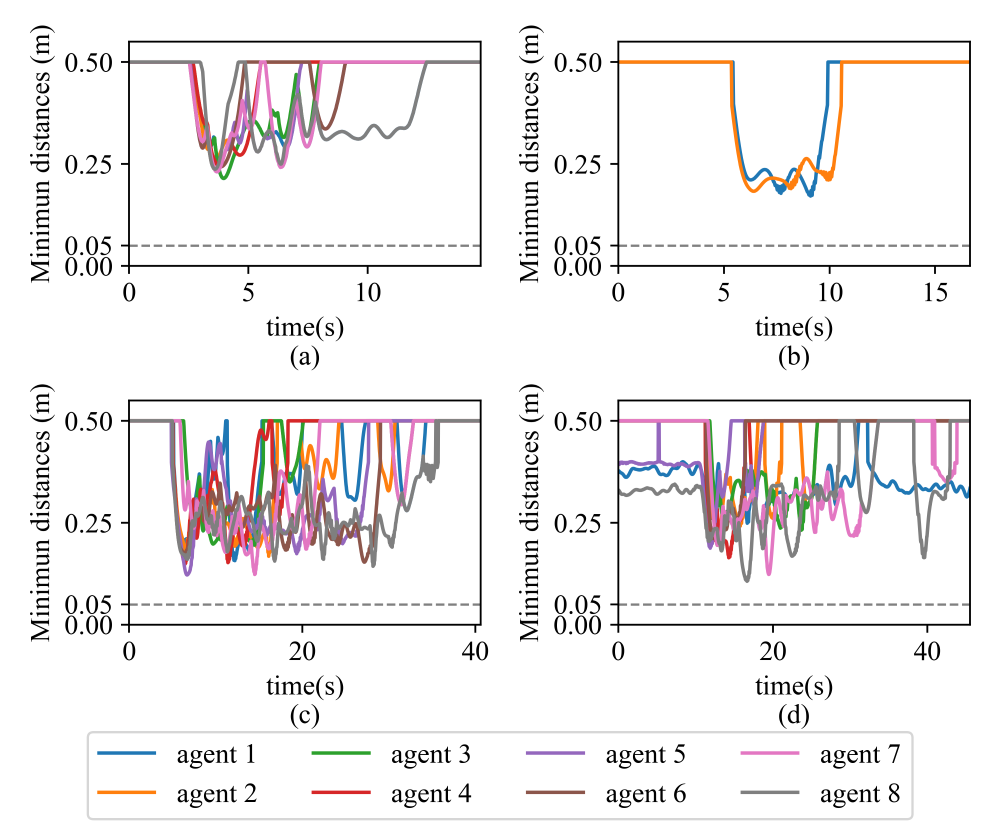}
    \caption{Hardware experiment results: Minimum distances between each agent and other agents or obstacles over time for four sets of experiments under $r=0.05{\rm m}$, corresponding to scenarios (a), (f), (k) in Fig.~\ref{fig.8} and (a) in Fig.~\ref{fig-mas}, respectively.}
    \label{fig.10}

    \vspace{3mm}

    \includegraphics[width=\linewidth]{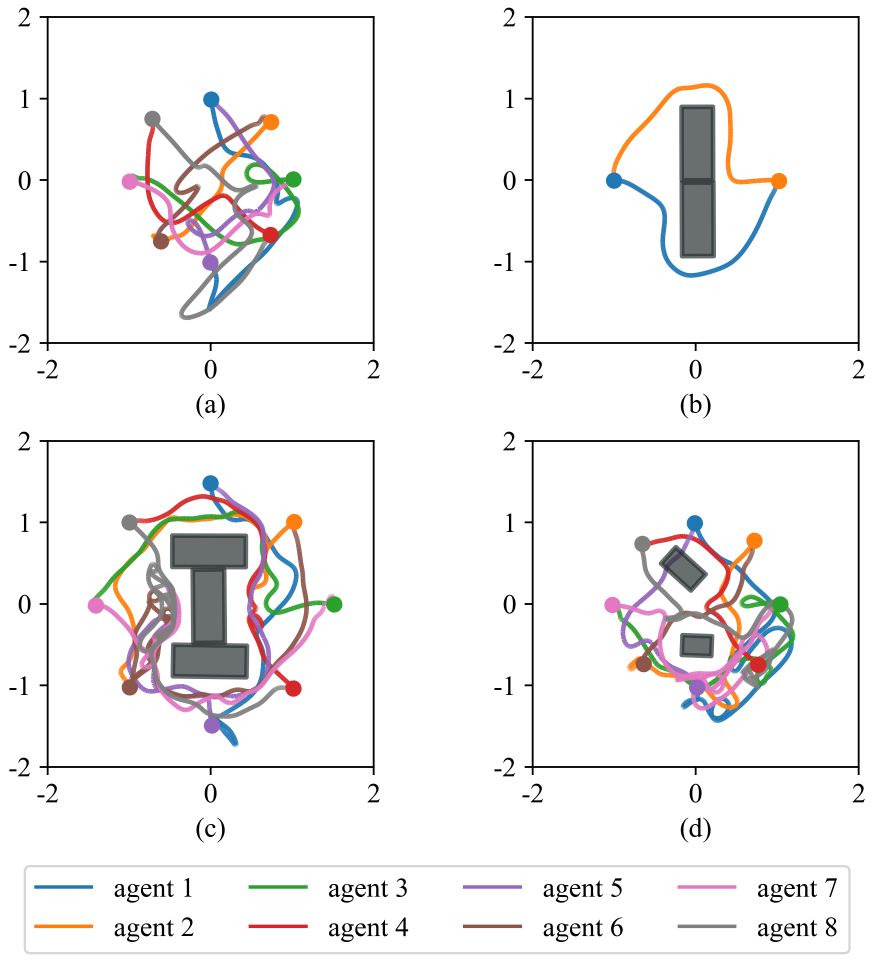}
    \caption{Simulation trajectories of all Crazyflie drones corresponding to four sets of experiments in Fig.~\ref{fig.8} and Fig.~\ref{fig-mas}. The upper gray rectangle in (d) indicates the initial position of the dynamic obstacle in Fig.~\ref{fig.8}(d).}
    \label{fig.11}
    \vspace{-5.5mm}
\end{figure}
Based on our HJB-GNN approach, the distributed safe policy is trained using a double integrator model, inspired by \cite{zhang2025tro}, to generate the desired accelerations. Similar to the experimental setup in \cite{zhang2025tro}, we utilize the \texttt{cmd\_full\_state} interface provided by Crazyswarm. The difference is due to the significant latency in Crazyflie’s onboard state estimation provided by Crazyswarm, we directly employ position and velocity measurements obtained from the OptiTrack system. 
To compute the \texttt{cmd\_full\_state} control command, we simulate the double integrator model used during training to predict the future positions and velocities resulting from the application of the desired accelerations after a fixed duration $\Delta t$. In all hardware experiments presented in this paper, we set $\Delta t = 50~\mathrm{ms}$. Each Crazyflie has a sensing radius of $R_a = 0.5\mathrm{m}$.

In addition, we present minimum distances between each agent and other agents or obstacles over time in Fig.~\ref{fig.10} and visualize Crazyflie drones' trajectories in Fig. \ref{fig.11} corresponding to four sets of experiments in Fig.~\ref{fig.8} and \ref{fig-mas}, respectively. It can be seen that all Crazyflie drones safely reach their respective goal positions.

\end{document}